\documentclass[aps,pra,twocolumn,groupedaddress,superscriptaddress,bibnotes,amsfonts,
citeautoscript,a4paper]{revtex4-2}
\usepackage{latexsym}
\usepackage{amsmath}
\usepackage{amssymb}
\usepackage{graphicx}
\usepackage[T1]{fontenc}
\usepackage[open]{bookmark}
\usepackage{hyperref}
\usepackage{orcidlink}
\usepackage{etoolbox}
\hypersetup{colorlinks=true,allcolors=blue}
\usepackage{lipsum} 
\newcommand{\vecnabla}{\bm{\nabla}}
\providecommand{\vect}[1]{{\mathbf{#1}}}
\DeclareRobustCommand{\uvec}[1]{{
  \ifcsname uvec#1\endcsname
     \csname uvec#1\endcsname
   \else
    \bm{\hat{\mathbf{#1}}}%
   \fi
}}

\newcommand{\wz}[1]{ {\color{cyan} Wenhua: #1} }

\newcommand{\comment}[1]{}
\def\ee{{\rm e}}
\def\ii{{\rm i}}
\def\rb{{\bf r}}
\def\Rb{{\bf R}}
\def\Eb{{\bf E}} \def\vb{{\bf v}}
\def\EELS{{\rm EELS}}
\def\ind{{{\rm ind}}}
\newcommand{\real}[1] {\mathopen{}{\rm Re}\left\{#1\right\}\mathclose{}}
\newcommand{\imag}[1] {\mathopen{}{\rm Im}\left\{#1\right\}\mathclose{}}
\def\wp{\omega_{\rm p}}
\def\eps{\epsilon}
\def\epsm{\epsilon_{\rm m}}
\def\um{{\mathrm{\text\textmu m}}}

\begin{document}

\title{Real-time surface plasmon polariton propagation in silver nanowires \\ 
 }

\author{Wenhua~Zhao\,\orcidlink{0009-0004-5721-607X}}
\affiliation{Max-Born-Institut, 12489 Berlin, Germany}
\affiliation{Department of Physics, AG Theoretische Optik \& Photonik, Humboldt-Universität zu Berlin, 12489 Berlin, Germany}
\author{Álvaro~Rodríguez~Echarri\,\orcidlink{0000-0003-4634-985X}}
\affiliation{Max-Born-Institut, 12489 Berlin, Germany}
%
\author{Alberto~Eljarrat\,\orcidlink{0000-0002-0968-5195}}
\affiliation{Department of Physics, Structure Research and Electron Microscopy group, Humboldt-Universität zu Berlin, 12489 Berlin, Germany}
\affiliation{Center for the Science of Materials Berlin, Humboldt-Universität zu Berlin, 12489 Berlin, Germany}
\author{Hannah~C.~Nerl\,\orcidlink{0000-0003-4814-7362}}
\affiliation{Department of Physics, Structure Research and Electron Microscopy group, Humboldt-Universität zu Berlin, 12489 Berlin, Germany}
\author{Thomas~Kiel\,\orcidlink{0000-0002-6070-9359}}
\affiliation{Department of Physics, AG Theoretische Optik \& Photonik, Humboldt-Universität zu Berlin, 12489 Berlin, Germany}
\author{Benedikt~Haas\,\orcidlink{0000-0002-9301-8511}}
\affiliation{Department of Physics, Structure Research and Electron Microscopy group, Humboldt-Universität zu Berlin, 12489 Berlin, Germany}
\affiliation{Center for the Science of Materials Berlin, Humboldt-Universität zu Berlin, 12489 Berlin, Germany}
\author{Henry~Halim\,}
\affiliation{Helmholtz-Zentrum Berlin for Materials and Energy,
14109 Berlin, Germany}
\author{Yan~Lu\,}
\affiliation{Helmholtz-Zentrum Berlin for Materials and Energy,
14109 Berlin, Germany}
\affiliation{Friedrich-Schiller-University, 07737 Jena, Germany}
\author{Kurt~Busch\,\orcidlink{0000-0003-0076-8522}}
\email{Corresponding author: kurt.busch@physik.hu-berlin.de}
\affiliation{Department of Physics, AG Theoretische Optik \& Photonik, Humboldt-Universität zu Berlin, 12489 Berlin, Germany}
\affiliation{Max-Born-Institut, 12489 Berlin, Germany}
\author{Christoph~T.~Koch\,\orcidlink{0000-0002-3984-1523}}
\email{Corresponding author: Christoph.Koch@hu-berlin.de}
\affiliation{Department of Physics, Structure Research and Electron Microscopy group, Humboldt-Universität zu Berlin, 12489 Berlin, Germany}
\affiliation{Center for the Science of Materials Berlin, Humboldt-Universität zu Berlin, 12489 Berlin, Germany}
\textbf{}

\date{\today}

\begin{abstract}

Electron microscopy techniques such as electron energy-loss spectroscopy (EELS) facilitate the spatio-spectral characterization of plasmonic nanostructures. In this work, a time-dependent perspective is presented, which significantly enhances the utility of EELS.
Specifically, silver nanowires offer the material and geometric features for various high-quality plasmonic excitations. This provides an ideal illustrative system for combined experimental-theoretical analyses of the different plasmonic excitations and their real-time dynamics. It is demonstrated how the plasmonic excitations propagating inside the wire repeatedly interact with the swift electrons in an EELS configuration. In addition, the role of azimuthal modes, often overlooked for very thin wires, is observed and analyzed in both the energy-loss spectrum and the dynamical perspective.
Such a complete understanding of the interaction of electrons and plasmonic excitation is key for the design of efficient plasmonic sensors, the study of hot electron dynamics in metals, and applications in the context of electron quantum optics, where full control of the spatial and temporal characteristics of the fields at the nanometer and femtosecond scales is highly desirable.




\end{abstract}

\maketitle

\section{Introduction}

The understanding of light-matter interaction at the nanoscale is a fundamental area of research that drives advances in a wide range of applications. In this context, the collective oscillations of conduction electrons in metals, the so-called plasmons (volume plasmons or, when spatially constrained by a material interface, surface plasmons), have proven to exhibit extraordinary properties that can be employed to boost light-matter interaction \cite{KZ12,xu2013near,sharma2012sers} for applications in circuitry \cite{bozhevolnyi2008plasmonic,Yu2019}, sensing \cite{AHL08, sensor_Arcadio, sensor_Jo}, quantum optics \cite{liu2017strong,Gonzalez2024}, quantum information \cite{slussarenko2019photonic}, and nonlinear optics \cite{SK16}. This is facilitated via the formation of hybrid light-matter excitations, the so-called plasmon polaritons, which come in the form of surface waves that propagate at dielectric-metal interfaces (surface plasmon polaritons) and spatially localized resonances (particle plasmon polaritons) \cite{Barnes-2003}. Amongst the commonly used plasmonic materials, silver provides the lowest intrinsic losses. This leads to higher-quality factors of resonances, and high-quality silver nanowires facilitate the propagation of surface plasmon polaritons for long distances so that silver is exceptionally suited for plasmonic applications \cite{paper258,RZK19,baburin2019silver}.

To excite and study plasmonic excitations, the use of swift electrons as in electron energy-loss spectroscopy (EELS) \cite{egerton2011electron} or cathodoluminescence (CL) \cite{liu2023modulation} allows probing plasmonic resonances at the nanometer scale while simultaneously retaining meV energy resolution \cite{paper085,bosman2007mapping,CKS16}. As a matter of fact, EELS and CL have successfully been employed to study plasmons in metals for several decades \cite{ruthemann1948diskrete,vincent1973dispersion,watanabe1956experimental,rossouw2011multipolar,rossouw2013plasmonic,zhou2014effect}. 

Both experimentally and theoretical/numerical analysis conventionally used in EELS of plasmonic systems, only provides information about the spectral position and localization in real space of resonances but no information about the actual time-dependent interactions between the electron and the plasmonic field.  However, this missing temporal component is essential for developing a more intuitive and complete understanding of the energy loss and gain mechanisms at play. Indeed, the interactions that occur between the sample and the electron beam while the electron approaches, passes by, or traverses the sample ultimately lead to the overall energy loss that the electron beam undergoes, yielding the resulting spectra. A complete understanding of such interactions and their dynamics is crucial to comprehend the underlying physical processes.
       

Here, based on the well-established EELS theory \cite{R1957,E96,paper149,Silcox1975}, we extract the temporal information out of conventional experimental EELS measurements by applying a Fourier transform (FT) to the loss spectrum of long silver nanowires in order to obtain a plasmonic excitation probability in the time domain. This study relies on the combination of the unprecedented high-energy resolution of modern electron microscopes and a highly efficient time-domain simulation approach to solve the full electromagnetic problem. Thereby, we characterize the dynamics of the corresponding plasmonic excitations. Moreover, we provide a comprehensive approach for the description of propagating surface plasmon
polaritons (SPPs) in long nanowires as well as a complete understanding of the above analysis of the conventional EELS measurements. Finally, we reveal the role of the often neglected azimuthal modes (higher-order SPPs) in the obtained EELS signals. 

\begin{figure}
    \centering
    \includegraphics[trim = 25mm 35mm 135mm 56.4mm, clip, width=0.5
    \textwidth]{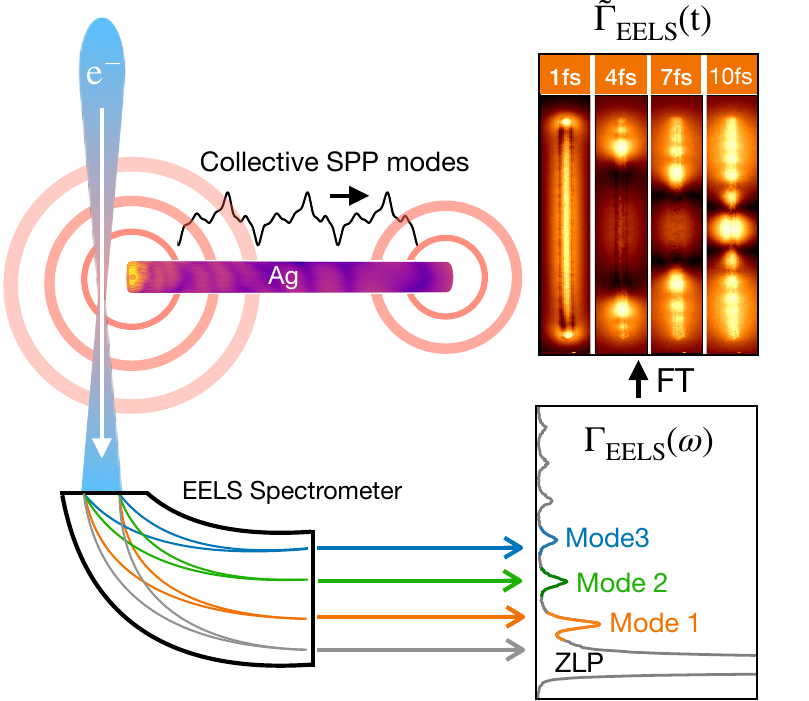}
    \caption{\textbf{From the electron energy loss to the plasmonic excitation probability.} 
    Illustration of electron energy-loss spectroscopy in a scanning TEM configuration. While the electron beam passes near the plasmonic silver nanowire, it excites surface plasmon polaritons (SPPs) that propagate from the nearest excitation point to the rest of the structure (see black curves manifesting the collective SPP modes, associated with the amplitude of the electric field shown by the colored map inside the Ag wire). Based on the resulting interaction of the induced plasmonic fields and the moving electron, the latter loses energy. Inside the spectrometer, these electrons are deflected according to their energy, which results in a loss spectrum showing peaks of intensity counts for different SPP modes. Through a Fourier transform (FT), we obtain the time evolution of the energy loss $\rm \tilde \Gamma_{\rm EELS}(t)$, where four snapshots display its magnitude at different times ($t=$ 1~fs, 4~fs, 7~fs, and 10~fs, defining $t=0$ when the electron is at the shortest distance from the nanowire) of the region of interest containing the nanowire.
}
\label{fig:Figure1}
\end{figure}

\begin{figure*}
    \centering
    \includegraphics[trim = 10mm 33mm 0mm 20mm, clip, width=1
    \textwidth]{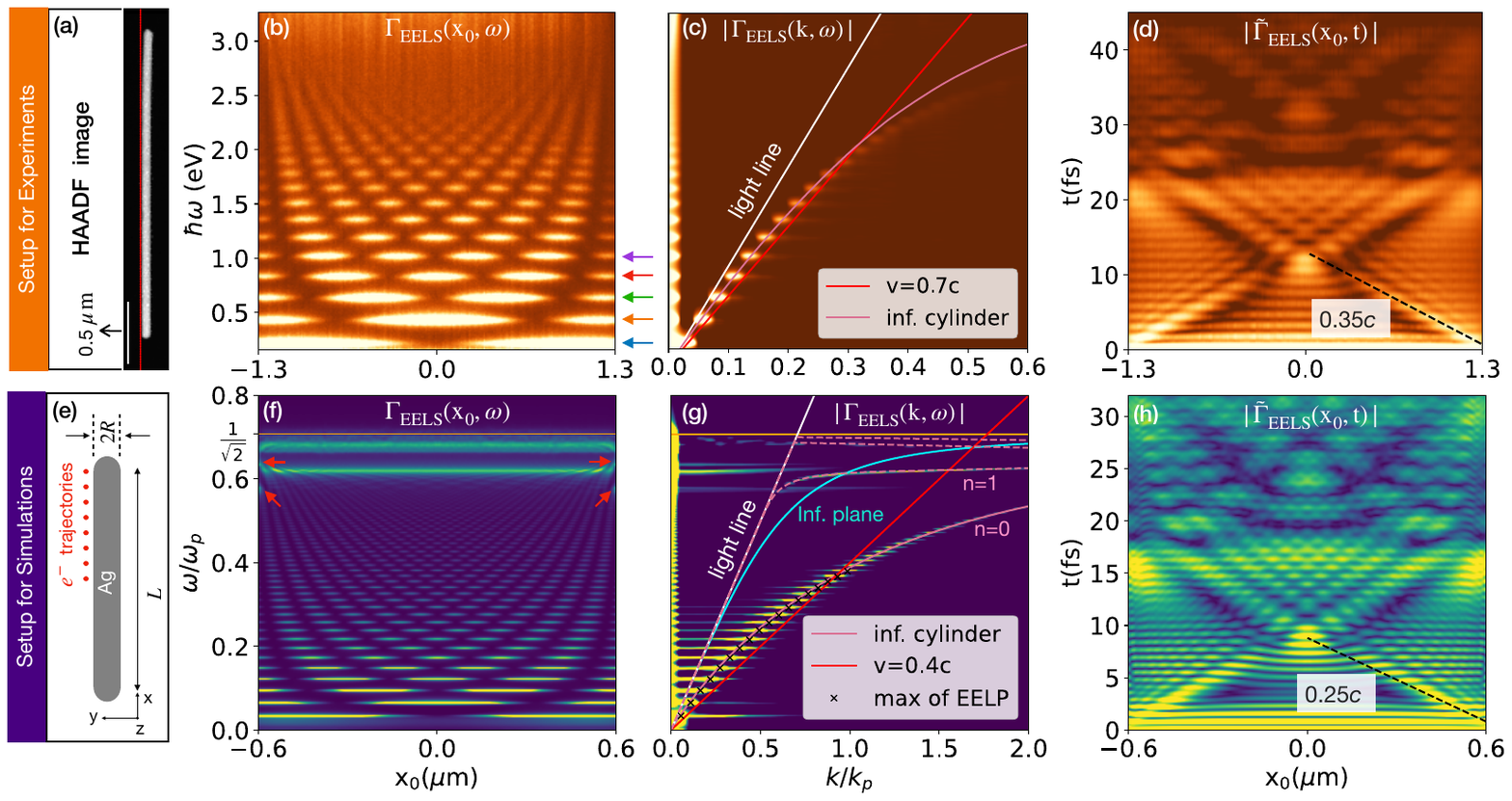}
    \caption{\textbf{Spatio-temporal characterization of plasmonic excitations in silver nanowires induced by a beam of swift electrons: Experiments and computations.} 
    \textbf{a} (experiments, always shown in red-yellow color coding) High-angle annular dark field (HAADF) image of a silver nanowire of length $\sim 2.6$~$\um$ and radius of $\sim 30$~nm.
    \textbf{b} (experiments) EELS line scan showing electron energy loss (vertical axis) as a function of position (horizontal axis) when acquired along the red line in panel (a), with aloof electron beams $\sim 5$~nm away from the surface of the wire. A $200$~keV electron beam energy ($v \approx 0.7 c$) is used.
    \textbf{c} (experiments) Dispersion relation from a spatial Fourier transform of (b) along the $x$ axis, together with that of an infinite silver cylinder of radius $30$~nm using measured optical data (pink curve) \cite{JC1972}, the electron line dispersion (red), and the light line (white) in units of $k_{\rm p} = \omega_{\rm p}/c$, with $\hbar\wp = 9.17$~eV the plasma energy of silver using Drude model.
    \textbf{d} (experiments) $|\tilde\Gamma_\EELS(t)|$ represents the temporal Fourier transform of the panel (b), i.e. the temporal evolution of the EELS signal for each electron trajectory. The slope of the feature moving towards the center of the panel propagates at velocity $\sim 0.35c$ (black dashed line).
    \textbf{e} (theory, always shown in yellow-blue color coding) Sketch of the system for numerical computations, where a free-standing silver nanowire of length $L=1.2$~$\um$ and radius of $R=15$~nm has been considered. The trajectory positions of the electrons in the $xy$-plane are marked as red dots, where the electrons travel in the $z$ direction. A $47$~keV electron beam energy ($v \approx 0.4 c$) has been utilized.
    \textbf{f} (theory) Spatially resolved EELS for the system shown in panel (e). The orange line gives the resonance frequency of the surface plasmon $\omega/\omega_{\rm p}=1/\sqrt{2}$.
    \textbf{g} (theory) Dispersion relation from a spatial Fourier transform of (f) along the $x$ axis.
    \textbf{h} (theory) Amplitude of $\tilde\Gamma_\EELS(t)$, the temporal Fourier transform of panel (f). The main feature traveling from the tip to the center shows a slope $\sim 0.25c$ (black dashed line). The simulation results for a wire of radius $R=30$~nm are displayed in Fig.~S3 in the SI. For simplicity, the color bars have been omitted. In all graphs, dark colors represent low values, while light colors represent high values (all graphs are shown on a linear scale).
    }
\label{fig:Figure2}
\end{figure*}


\section{Results and discussion}
In our experimental setup (see Fig.~\ref{fig:Figure1}), we consider a swift electron interacting with a long silver nanowire with a few microns in length and a few tens of nanometers in diameter, which was placed on a thin silicon nitride substrate. We acquired the corresponding EELS maps using a state-of-the-art scanning transmission electron microscope (STEM), which provides high-resolution spectral information (for details, see Sec.~\ref{Sec:experiments_methods} and Sec.~\ref{Sec:microscopyMethods} in the Methods section).

To complement the experimental observations, we have performed numerical computations of the system by solving Maxwell's equations using the Discontinuous-Galerkin Time-Domain (DGTD) finite-element method \cite{DGTD,Elli_Wenhua2023,Kiel_metalspitze}, which facilitates the
spatio-temporal resolution of the plasmonic modes in the nanometer and femtosecond range (for details, see Sec.~\ref{Sec:DGTD_methods} in
the Methods section). Importantly, we modeled the material properties of a free-standing silver nanowire of radius $R$ and length $L$ using a local Drude permittivity $\epsm = \eps_0-\wp^2/\omega(\omega+\ii\gamma)$ with bulk plasmon energy $\hbar\wp = 9.17$~eV, damping $\hbar\gamma = 21$~meV, and $\eps_0=1$ \cite{JC1972} [see Sec.~S4 in the Supplementary Information (SI) for further insights about the influence of interband transitions].

\subsection{Temporal electron energy-loss spectroscopy}

The energy loss experienced by a swift electron (moving at velocity $\vb$) originates from the backaction of the induced fields on the electron and reads \cite{R1957,paper149}
\begin{equation} \label{eq:energyLoss}
    \Delta E = e \int \mathrm{d}t \, \vb \cdot \Eb^\ind[\rb_e(t),t] = \int_0^\infty \mathrm{d}\omega \, \hbar\omega \, \Gamma_\EELS(\omega).
\end{equation}
Here, $-e<0$ is the electron charge. Assuming that the electron is moving in the $z$ direction, the position of the electron at time $t$ is $\rb_e(t)=(x_0, y_0, z_0+vt)$, where $(x_0,y_0)$ denotes the electron trajectory position 
on the $xy$ plane. Further, for aloof trajectories the impact parameter 
$b\equiv |y_0|-R$ gives the shortest distance between the electron trajectory and the surface of the wire. In the no-recoil approximation \cite{paper149}, we then obtain
\begin{equation} \label{eq:EELP_dw}
    \Gamma_\EELS (\omega) = \frac{1}{\pi \hbar\omega} \int_{-\infty}^{\infty} \mathrm{d}z \; \real{ E^{\rm ind}_z(z,\omega) \, \ee^{-\ii \omega \frac{z-z_0}{v}}},
\end{equation}
which is the so-called loss probability in units of transferred frequency $\omega>0$ (i.e., associated with an energy loss $\hbar \omega$). Note 
that the induced electric field $E^{\rm ind}_z$ refers to the $z$ component (i.e., parallel to the electron trajectory), and the integral extends over 
the electron trajectory. We define its Fourier transform via
\begin{flalign}  \label{eq:EELS_tilde_time}
    \tilde\Gamma_\EELS(t) = \int_{-\infty}^{\infty} \frac{\mathrm{d}\omega}{2\pi}\, \Gamma_\EELS(\omega) \, \ee^{-\ii \omega t}.
\end{flalign}
By considering the imaginary part of $\tilde\Gamma_\EELS(z, t)$, we obtain
\begin{flalign}  \label{eq:EELS_tilde_time_im}
    {\rm Im}\{\tilde\Gamma_\EELS(t)\} = -\frac{e}{4\pi \hbar} \; \int_{-\infty}^{\infty} \mathrm{d}z \; \tilde\Gamma_\EELS(z, t), 
\end{flalign}
where $\tilde\Gamma_\EELS(z, t) = \int_{t'-t}^{t'+t} \mathrm{d}\tau \; E_z^\ind(z, \tau)$. Further, $t'=(z-z_0)/v$ denotes the arrival time of 
the electron at position $z$. Instead of integrating the electric field along the position of the electron as in Eq.~\eqref{eq:EELP_dw}, in this expression, for each position $z$ we integrate over a temporal window of width $2t$ centered around the arrival time $t'$  (see details in Sec.~S1 in the SI). At time $t=0^{+}$, we obtain $\tilde\Gamma_\EELS(z, 0^{+}) \propto E_z^\ind(z, t')$, which is the induced electric field at the position of the electron. 
%
\comment{
    \begin{flalign} \label{eq:Gamma_t_Ez}
    \tilde\Gamma_\EELS(t) &\approx \frac{\mathrm{1}}{2\pi \ii} \; \int_{t'-t}^{t'+t} \mathrm{d}\tau \; E_z^\ind[z_{\rm sur}, \tau] \\ \nonumber
    &\approx \frac{\mathrm{1}}{2\pi \ii} \; \int_{-\infty}^{t'+t} \mathrm{d}\tau \; E_z^\ind[z_{\rm sur}, \tau] \\ \nonumber 
    &\approx \frac{\mathrm{1}}{2\pi \ii} \; \int \mathrm{d}t \; E_z^\ind[z_{\rm sur}, t],
\end{flalign}
\wz{the last step is probably not quite correct ....}
where the approximation in the first step is based on the fact that $E_z^\ind[z_{\rm sur}, \tau]$ at the position $z_{\rm sur}$ near the surface of the wire contributes the most to the integral over $z$. For the approximation in the second step, we utilize the fact that the electric field amplitude for the time before the electron meets the nanowire is nearly zero. Note that in the third step, we set the arrival time of the electron at $z_{\rm sur}$ as 0. Next, we connect $\tilde\Gamma_\EELS(t)$ with the induced electric field $E_z^\ind[z_{\rm sur}, t]$ at ($\Rb_0, z_{\rm sur}$) near the surface of the nanowire on the trajectory. Thus we associate $\tilde\Gamma_\EELS(t)$ with the return time of the plasmon modes that bounce at the end of the nanowire and reach ($\Rb_0, z_{\rm sur}$). More details are given later in the manuscript. 
}

%
As demonstrated in Fig.~\ref{fig:Figure1}, $\tilde\Gamma_\EELS(t)$ represents a dimensionless quantity providing us with information about the excitation probability of the plasmonic modes in the nanowire. There, we depict the amplitude of $\tilde\Gamma_\EELS(t)$ as a function of space and time, thereby creating characteristic maps of the excitation probability. Despite the electron being detected far from the interaction region, the information we obtain is directly related to the electric near-field distribution \cite{Polman_2019}, as we will demonstrate below. While we focused on the FT of the EELS signal, further analyses in the time domain could be performed. For instance, we could consider the time derivative of $\tilde\Gamma_\EELS(t)$, corresponding to the Fourier transform of $\hbar \omega \Gamma(\omega)$ in the time domain. Consequently, various alternative ways exist to extract information from the time domain.

Due to their reduced dimensionality and low material losses in conjunction with smooth surfaces, chemically prepared silver nanowires are the perfect venue to study propagating SPPs. Specifically, for a wire $L\approx 2.6$~$\um$ long and $R\approx30$~nm radius, as shown in the high-angle annular dark field (HAADF) image \cite{TEM} in Fig.~\ref{fig:Figure2}a, we collected the electron-energy loss (EEL) spectra along the red line ($b\approx 5$~nm) to obtain the line scan in Fig.~\ref{fig:Figure2}b, for a $200$~keV electron beam energy ($v \approx 0.7 c$). To assess the role of electron velocity experimentally, the measurements were repeated with 30~keV and 60~keV electron acceleration energies, and the outcomes were found to be directly comparable. 
The spatial variations, as observed in the experimental EELS map result from the interaction of the electron beam with the spatial profiles of the different plasmonic modes. These modes resemble standing waves bounded by the caps of the wire. In order to obtain the dispersion relation of the plasmons (i.e., $\hbar \omega$ energy and $k$ momentum diagram) as displayed in Fig.~\ref{fig:Figure2}c, we computed the spatial Fourier transform of the EELS maps and compared with the analytic dispersion relation for an infinite cylinder (pink curve) for the fundamental angular modes $n=0$ \cite{AE1974} using measured optical data for silver \cite{JC1972}. More generally, an angular mode with mode index $n$ exhibits a spatial profile of the form $\ee^{\ii n \varphi}$ along the azimuthal direction $\varphi$. Both the experimental and analytical dispersion relations agree well so that the experimentally measured wire can be considered to be sufficiently long so that its dispersion relation depends only on the radius and the material properties \cite{AE1974} (see Sec.~S2 in the SI). In addition, we calculate the excitation probability map in time as shown in Fig.~\ref{fig:Figure2}d, following Eq.~\eqref{eq:EELS_tilde_time} (i.e., the temporal Fourier transform of Fig.~\ref{fig:Figure2}b). Note that, while in Fig.~\ref{fig:Figure1}, we have shown snapshots of the entire region of interest (in the $xy$ plane), in Fig.~\ref{fig:Figure2}d, we show the results corresponding to the electron beam positions indicated by the red line in Fig.~\ref{fig:Figure2}a. The full video in the $xy$ plane ["Movie$\_$EXP.mp4"] is shown in the Supplementary Material.

In order to obtain insight into the excitation probability maps displayed in Fig.~\ref{fig:Figure2}, we computed the EELS for a long silver nanowire with length $L = 1.2$~$\um$ and radius $R=15$~nm numerically via the DGTD solver using Eq.~\eqref{eq:EELP_dw} for the trajectories indicated by the red dots (see Fig.~\ref{fig:Figure2}e) and show the results in Fig.~\ref{fig:Figure2}f. In these computations, we used a shorter wire than in the experiment, since we found it to be sufficiently long to obtain a well-defined dispersion relation, and an electron velocity $v = 0.4 c$ ($\approx 47$~keV). When comparing experiments and simulations, we found that the EELS maps are rather similar (c.f., Fig.~\ref{fig:Figure2}b and Fig.~\ref{fig:Figure2}f, respectively). 
By performing a spatial Fourier transform of the measured and simulated EELS maps along the $x$ axis, we obtained the dispersion relation in Fig.~\ref{fig:Figure2}c and Fig.~\ref{fig:Figure2}g, respectively. Note that the longer wire in the experiment leads to a better contrast in the experimental dispersion features relative to those in the computations as the FT of a longer system in real space leads to more localized features in the reciprocal space.
We further notice that the experimental dispersion matches well with that of an infinite silver cylinder using measured optical data (pink curve), as shown in Fig.~\ref{fig:Figure2}c. Similarly, the computed dispersion of the SPPs matches the dispersion of an infinite cylinder (pink curve) calculated using a local Drude model, as expected for a sufficiently long nanowire. Since the radius of the wire is small ($R=15$~nm), the dispersion curve is far from the dispersion relation of a planar silver/air interface (blue curve). As a matter of fact, the latter case corresponds to the limiting situation when the radius of the cylinder is infinite, or when the characteristic size of the excitations is much smaller than the curvature of the structure. Therefore, in the $Rk\to\infty$ limit, all lines converge to the value $\omega/\wp = 1/\sqrt{2}$, at least when neglecting the screening effects of interband transitions in the material properties. Further, the higher-order mode with $n=1$ is also visible in the dispersion relation in Fig.~\ref{fig:Figure2}g. First, we focus on the fundamental modes ($n=0$) that correspond to the standing-wave-like modes along the wire with very well-defined energy and parallel momentum (see black crosses indicating the maxima of the colormap in Fig.~\ref{fig:Figure2}g). Below, we address the effect of the azimuthal modes ($n\ge1$). Additionally, at energy $\omega/\wp \sim 0.6$ (shown in Fig.~\ref{fig:Figure2}f) for trajectories close to the termination of the wire, there are further features in the EEL spectra (indicated by red arrows), which we can be attributed to the geometrical shape of the caps and the fact that despite being a long wire it is still finite in length. In our case, the terminations of the wire were simulated with hemispherical caps, and the bright horizontal lines around the $n=1$ mode in the dispersion relation of Fig.~\ref{fig:Figure2}g are related to the localized Mie resonances at these caps. For further clarification, we have superimposed Mie calculations using the MNPBEM toolbox \cite{MNPBEM} for a full sphere to the EELS map, and this corroborated the origin of these features (see also Sec.~3 in the SI).

Furthermore, there is a very good qualitative agreement between the measured and computed excitation probabilities, see Figs.~\ref{fig:Figure2}d and ~\ref{fig:Figure2}h. However, the values of the slopes characterizing the main features differ because they have different radii, and this leads to different dispersion relations. Consequently, we repeated the simulations for a wire with a larger radius $R=30$~nm (shown in Fig.~S4 in the SI). In this case, the computed values of the slope do match the experimentally obtained one. Indeed, the fitting of the feature is more easily visible when looking at the imaginary part of $\tilde\Gamma_\EELS(t)$ (see Fig.~S2 in the SI, where we show the real and imaginary parts). Therefore, we use the $\imag{\tilde\Gamma_\EELS(t)}$ below to illustrate the excitation probability dynamics. This excellent agreement between experiments and computations further highlights the validity of our approach of using the simulation tools in the time domain to analyze the resulting excitation probability. 

\begin{figure}[t]
    \centering
    \includegraphics[trim = 70mm 6.5mm 70mm 36mm, clip,width=0.5
    \textwidth]{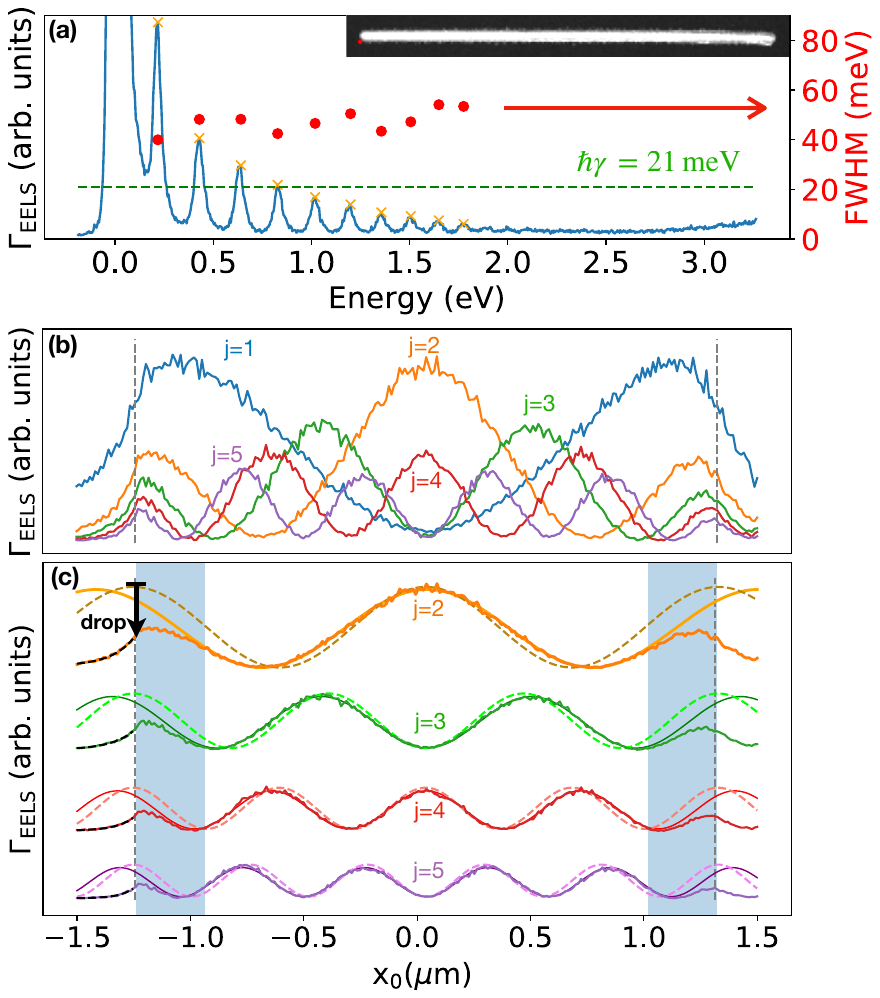}
    \caption{\textbf{Effect of the caps on the EEL spectra of silver nanowires (experiments).} 
    \textbf{a} Left vertical axis: Experimental EEL spectrum of $2.6$~$\um$-long silver wire with $30$~nm radius excited with a
    200~keV electron beam passing aloof near the cap (see image in the inset, where the red dot indicates the position of the electron trajectory). Right axis: Full-width at half maximum (FWHM) for each of the peaks (yellow crosses) and comparison with the Drude 
    damping rate of 21~meV (green dashed line). 
    \textbf{b} Intensity profiles of the EELS peaks for trajectories along the red line in Fig.~\ref{fig:Figure2}a ($b\approx5$~nm) for different plasmonic modes in Fig.~\ref{fig:Figure2}a (see colored arrows in Fig.~\ref{fig:Figure2}b). The two vertical grey dashed lines mark the caps of the nanowire. 
    \textbf{c} Same as (b) but vertically offset and fitted using cosine functions. Bold solid curves show the fits to experimental data, limited to within the central white region to remove the effect of the caps (blue regions). The colored dashed lines show the analytical model: $\cos^2(k_j x)$ using $k_j=j\pi/L$. The black dashed lines outside of the nanowire result from fits of exponential functions corresponding to the decay of the signal for the region outside. For each of the modes, we define a \emph{drop} (black arrow) between the analytical model and the experimental EELS.
    }
\label{fig:Figure3}
\end{figure}

\begin{figure*}[t]
    \centering
    \includegraphics[trim = 0mm 26.9mm 0mm 40mm, clip, width=1
    \textwidth]{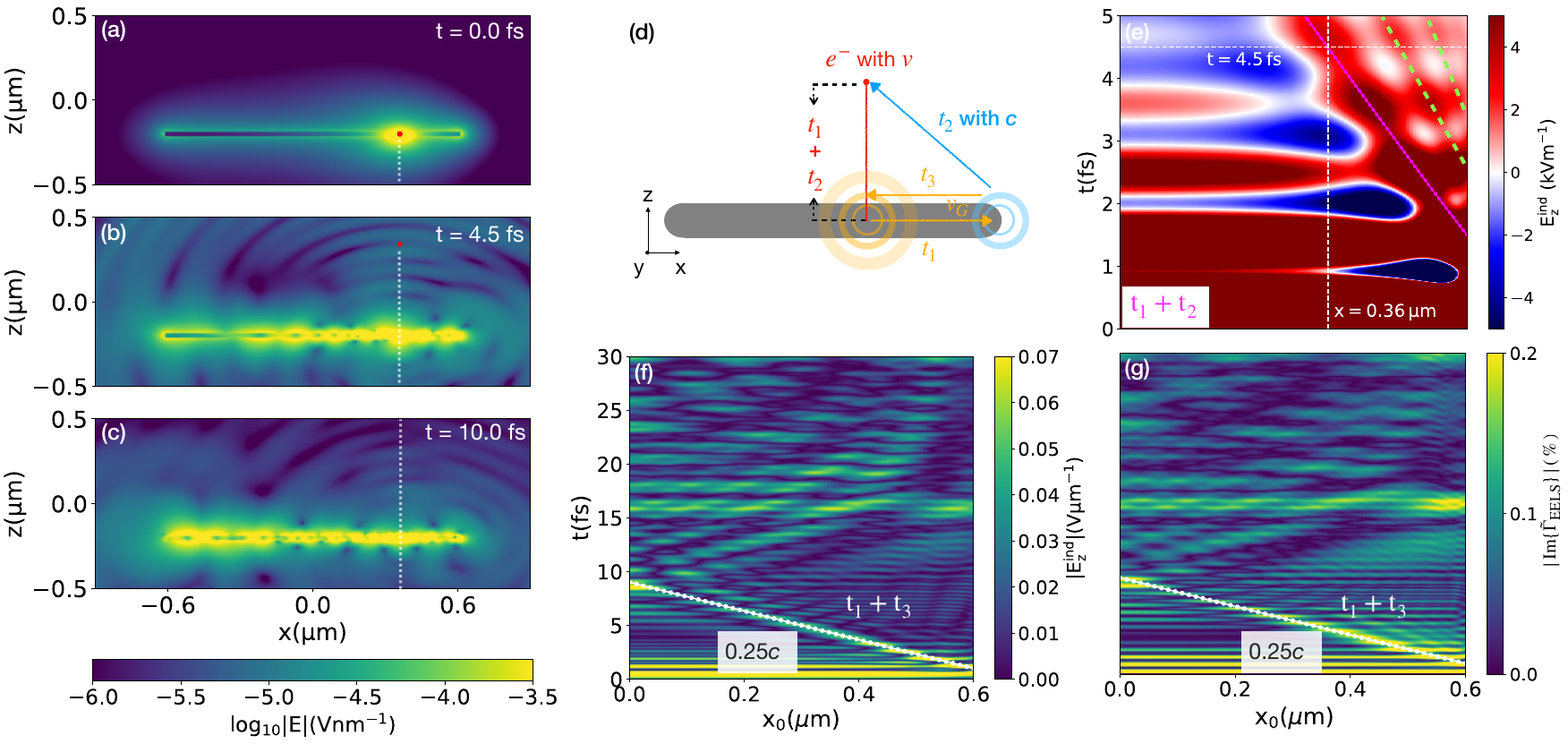}
    \caption{\textbf{Theoretical interpretation of the temporal EELS signal.} 
    Snapshots of the induced electric fields for an electron at different positions: \textbf{a} right after arriving at the silver nanowire of $R=15$~nm and $L=1.2$~$\um$, \textbf{b} after passing the wire, and \textbf{c} leaving the computation domain (see exact times in insets). 
    \textbf{d} Schematic of the return time dynamics, see more in the main text.
    \textbf{e} Electric field at the position of the electron for different electron trajectories given in Fig.~\ref{fig:Figure2}e, together with return time $t_1+t_2$ (pink curve). 
    For the same system, we show in \textbf{f} the electric field $\Eb(\rb_s,t)$ near the surface of the nanowire (i.e., $\rb_s=[x_0, R+b, z_s]$) for different trajectories, where $R<z_s<2\pi c/\wp$ within the near-field region. The white dotted line indicates the SPP round-trip time $t_1+t_3$ using the group velocity $v_G \approx 0.5c$.
    \textbf{g} Similar as in Fig.~\ref{fig:Figure2}f, but for $|{\rm Im} \{\tilde\Gamma_\EELS(t)\}|$, together with SPP round-trip time $t_1+t_3$ (dotted white line).
}
\label{fig:Figure4}
\end{figure*}

\subsection{EELS: The electric field at the position of the electron}

The excitation probability maps displayed in Figs.~\ref{fig:Figure2}d and ~\ref{fig:Figure2}h were obtained along a line at a distance of $b=5$~nm from the cylinder, while in Fig.~\ref{fig:Figure1}, we show the experimentally acquired excitation probability map in the $xy$ plane for specific times (see labels). Interestingly, the electron moves at the speed of $v=0.4c$ (in the numerical simulation and even faster in the measured data), meaning that it passes by the 30~nm-thick wire at a minimum distance given by the impact parameter ($5$~nm) in about a tenth of a femtosecond. For convenience, we set the reference $t=0$ as that point in time in which the electron \emph{arrives} at the wire. In Fig.~\ref{fig:Figure1} and ~\ref{fig:Figure2}d, we observe how well-defined features with large excitation probability amplitudes are emanating from the caps (symmetrically from either side) to the center. Despite being a long wire, the fact that it is finite produces such enhancements in the electron losses. 

Inside the wire, there are two main loss mechanisms that are directly responsible for the electron energy loss observed: Ohmic losses associated with the imaginary part of the dielectric function of the metal and radiative losses. Analyses of the EELS peak widths (shown in Fig.~\ref{fig:Figure3}a) for experimentally measured spectra with an aloof trajectory near the wire cap (see the inset in Fig.~\ref{fig:Figure3}a) and for a simulated structure (see Fig.~S5d in the SI), clearly show that in both cases, the widths of all peaks are larger than the intrinsic Drude damping of $\hbar\gamma = 21$~meV (see green dashed line). Such an effect has previously been reported \cite{paper369}, and it is a tell-tale of the combination of radiative and non-radiative losses. 

As observed in Fig.~\ref{fig:Figure3}b, radiation losses occur due to the surface roughness of the wire and the scattering at the wire caps. There, we show the measured EELS along the red line in Fig.~\ref{fig:Figure2}a, where the middle point of the wire is set to $x_0=0$, and dashed vertical lines limit the extension of the wire at $\pm L/2$. Different modes $j$ (indicated by different colors and corresponding to those highlighted in Fig.~\ref{fig:Figure2}b by arrows) exhibit well-defined spatial profiles and manifest a partial drop in the EELS signal at the caps of the nanowire. To characterize an individual's mode drop, we plotted the analytical standing waves using $ \propto \cos^2(k_jx)$ with $k_j=j\pi/L$ (colored dashed lines) for selected modes $j$. For each mode, we estimate the drop in the EELS signal between the standing waves model and the experimental EELS (see black arrow in Fig.~\ref{fig:Figure3}c) to be 58.6\%, 73.3\%, 79.3\%, and 85.0\% for modes $j=2,\, 3,\, 4, \text{ and } 5$, respectively. We determined the individual modes' wave vectors by fitting the standing-wave model to the experimental EELS data, excluding the cap regions (i.e., considering the EELS within the white area in Fig.~\ref{fig:Figure3}c). The thus obtained the wave vectors $k_{j,\rm fitting}$ that compares with $k_j$ as $k_j/k_{j,\rm fitting}= 1.13$, 1.07, 1.06, and 1.05 for modes $j=2,\,..., \,5$, respectively. This indicates that the effective spatial extent of the modes is slightly larger than the actual size of the nanowire. Interestingly, the spatial profile of the modes resembles those of a Fabry-Pérot cavity \cite{SalehTeich_Fundamentals} (see Fig.~S10b in the SI, where the profile of the modes outside the cavity exponentially decays (see black dashed lines in the left-hand side of Fig.~\ref{fig:Figure3}c). Importantly, all modes have a partial drop in EELS intensity, indicating that part of the electromagnetic energy of the mode gets reflected and transmitted at the wire caps (as shown below).


 

One of the main advantages of the DGTD approach to solving Maxwell's equations is that we obtain the time-dependent induced fields, which allows for studying the dynamics of the overall system time step by time step. Figs.~\ref{fig:Figure4}a-c show the magnitude of the electric field $|\vect{E}|=\sqrt{E_x^2+E_y^2+E_z^2}$ in the $xz$-plane for different instances of time. In each graph, the position of the electron is indicated as a red point on the specific electron trajectory with $x_0=0.36$~$\um$ and impact parameter $b=5$~nm. The electron reaches the center of the nanowire around time $t\approx0$~fs, leading to large field values at the electron's position (see Fig.~\ref{fig:Figure4}a). Simultaneously, different SPP modes are excited. At a later time, see Fig.~\ref{fig:Figure4}b, the excited SPP wave packets have propagated away from the intersection of the electron trajectory and the wire to either side. In the meantime, the electron has moved further away from the nanowire and experiences the radiated field originating from the intersection (note the presence of wavefront fringes). For this specific example, the SPP wave packet that moves to the right reaches the cap of the nanowire
before the left-moving SPP wave packet. When this right-moving SPP wave packet reaches the cap, it is scattered, and an almost spherical radiation burst is emitted. This burst travels at the speed of light in vacuum so that parts of it catch up and interact with the slower-moving electron. Roughly, it is at time $t\approx4.5$~fs that the first wavefront of the radiation burst reaches the electron, which is clearly seen in the video ["Movie$\_$DGTD.mp4"] provided in the Supplementary Material. Later, when the electron has already left the computational domain, we observe that the SPP wave packet moving to the left reaches the left cap, again leading to the emission of a radiation burst which, relative to the first burst, is delayed by about $\sim 8$~fs. The resulting wave fronts become visible and interfere with the rest of the electromagnetic fields (Fig.~\ref{fig:Figure4}c). Through the snapshots of Figs.~\ref{fig:Figure4}a-c, we study the field strength that the electron experiences when traveling along its trajectory. Interestingly, the cap-related EELS effect is a consequence of the finite length of the wire, and the underlying mechanism is illustrated in Fig.~\ref{fig:Figure4}d. Such a scheme shows that when the electron meets the nanowire, the excited SPP wave packet moves with group velocity $v_G$ to the right cap during a time $t_1$, then the right cap starts to radiate. The resulting wavefronts travel at the speed of light $c$ in vacuum, and after a time $t_2$ they reach the electron. In parallel, the electron travels at $v=0.4c$ for a time $t_R=t_1+t_2$, which we define as \emph{return time}, which is the time it takes for the initial excitation at $t=0$ to reach the electron again. The return time is calculated as
\begin{equation} \label{eq:return_time}
    v^2(t_1+t_2+\Delta)^2 + t_1^2v_G^2 = t_2^2c^2.
\end{equation}
%
Here, $\Delta$ denotes a small delay between the arrival time of the electron at the center of the nanowire and the time at which the SPP wave packets start to propagate due to the retardation that is required to build up the material response. 

At first glance, one might assume that the group velocity $v_G$ of the excited SPP wave packets in the wire would be determined by the slope $\mathrm{d}\omega/\mathrm{d}k$ of the SPP dispersion relation at the intersection of the electron with SPP dispersion, i.e. the 
intersection between the red curve (describing the electron) and the SPP dispersion relation in Fig.~\ref{fig:Figure2}g. In this case, the group velocity's value would be $\mathrm{d}k/\mathrm{d}\omega = 0.26c$). Further, one might assume that, depending on the kinetic energy of the electron (i.e., its velocity), one could control the propagation velocity of the excited SPP wave packets. 
However, we have observed that this is not the case. To demonstrate this, we depict in Fig.~S6 in the SI the electric field $|\vect{E}|$ as a function of time along a line parallel to the nanowire in the $x$ direction when the electron travels $b=5$~nm away from the cap for different parameters (i.e., radius, electron kinetic energy, and length). The two vertical dashed white lines at $x=\pm 0.6$~$\um$ indicate the caps of the nanowire. Around $t\approx0$~fs, the electron arrives at the point of minimal distance to the nanowire, and we observed that a SPP wave packet starts to travel from the right to the left side with a group velocity of $v_G\approx0.5c$ as determined by fitting the slope (red dashed curve). We found the resulting velocity to be dependent on the radius of the wire, namely its dispersion relation. Hence, its value is independent of the nanowire length and the electron kinetic energy. Note that, due to the delayed material response and the finite impact parameter, the resulting SPP wave packet requires about $\Delta \sim 1$~fs to form. Moreover, after the main SPP wave packet has formed, additional smaller satellite packets moving with slower velocities, as well as features that move at the speed of light, emerge. The latter corresponds to the induced field moving in free space and not inside the wire). However, due to the interference of all these satellite features, a rather complex dynamics results. In other words, all SPP modes are excited at once and propagate together in a complex manner. Below, we will unravel the corresponding dynamics. 

Based on the fitted group velocity and using Eq.~\eqref{eq:return_time}, we depict in Fig.~\ref{fig:Figure4}e the electric field at the position of the electron for different trajectories (i.e., for different $x_0$ with fixed impact parameter $b=5$~nm), where the return time $t_R=t_1+t_2$ of the first bright SPP wave packet is superimposed (pink curve). We notice an excellent match between the strong enhancement of the electric field at the position of the electron and the return time, in agreement with the scheme of Fig.~\ref{fig:Figure4}d. The features of lower intensity, denoted with green dashed lines in the right top corner of Fig.~\ref{fig:Figure4}e, originate from the aforementioned SPP satellites. Note that for the trajectory $x_0=0.36$~$\um$, we also highlight the time $t=4.5$~fs (white dashed lines), corresponding to the time in Fig.~\ref{fig:Figure4}b, where we observe a strong enhancement of the electric field at the electron position. Hence, such a strong enhancement is due to the interaction of the scattered field of the propagating modes originating at the right tip of the wire. The graph in Fig.~\ref{fig:Figure4}f shows the electric field $E_z^{\rm ind} (\rb_s,t)$ at the position near the surface of the wire at different electron trajectories (i.e., $\rb_s=[x_0, R+b, z_s]$, where $R<z_s<2\pi c/\wp$). Complementary, we display in Fig.~\ref{fig:Figure4}g the map $|{\rm Im} \{\tilde\Gamma_\EELS(t)\}|$ along the $x$-axis for the same time axis as in Fig.~\ref{fig:Figure4}f. We notice that these two plots are very similar. This indicates that $\imag{\tilde\Gamma_\EELS(t)}$ is closely related to the near-field profile of the excited plasmonic modes given by $E_z^{\rm ind} (\rb_s,t)$. 

Here, we define the round-trip time $t_1+t_3 \approx 2 t_1 = 2 |L/2-x_0|/v_G$, which is the time the excited SPP wave packet requires to travel to the right cap and to return to its launch point $x_0$. Both, Fig.~\ref{fig:Figure4}f and Fig.~\ref{fig:Figure4}g, show the $t_1+t_3$ time for each value of $x_0$. We notice that the bright features in both plots match well with the round-trip time. Therefore, this theoretically identified close connection between $\tilde \Gamma_\EELS(t)$ and the round-trip time $t_1+t_3$ makes possible the calculation of $2 t_1$ return time directly from the experimental EELS data, thereby fixing the group velocity of the propagating SPP wave packet.

For the excitation probability in Fig.~\ref{fig:Figure2}d, we obtain the slope of the bright feature to be propagating at $0.35c$, indicating that the group velocity of the SPP wave packets in the experiments is $v_G\approx0.7c$, which matches the result from the simulation (for further details, see Fig.~S6d in the SI). Note that the maps $\tilde \Gamma_\EELS(t)$ in Fig.~\ref{fig:Figure4}f and Fig.~\ref{fig:Figure4}g do not change for different electron velocities since the dispersion relation of the plasmonic modes does not depend on the excitation. Interestingly, while the SPP profile remains qualitatively the same for different kinetic energies of the electron, the electric field profile at the electron position nonetheless changes (see Fig.~S8 in the SI).

\subsection{Propagating modes and density of optical states}

\begin{figure}
    \centering
    \includegraphics[trim = 45mm 9mm 35mm 0mm, clip, width=0.5\textwidth]{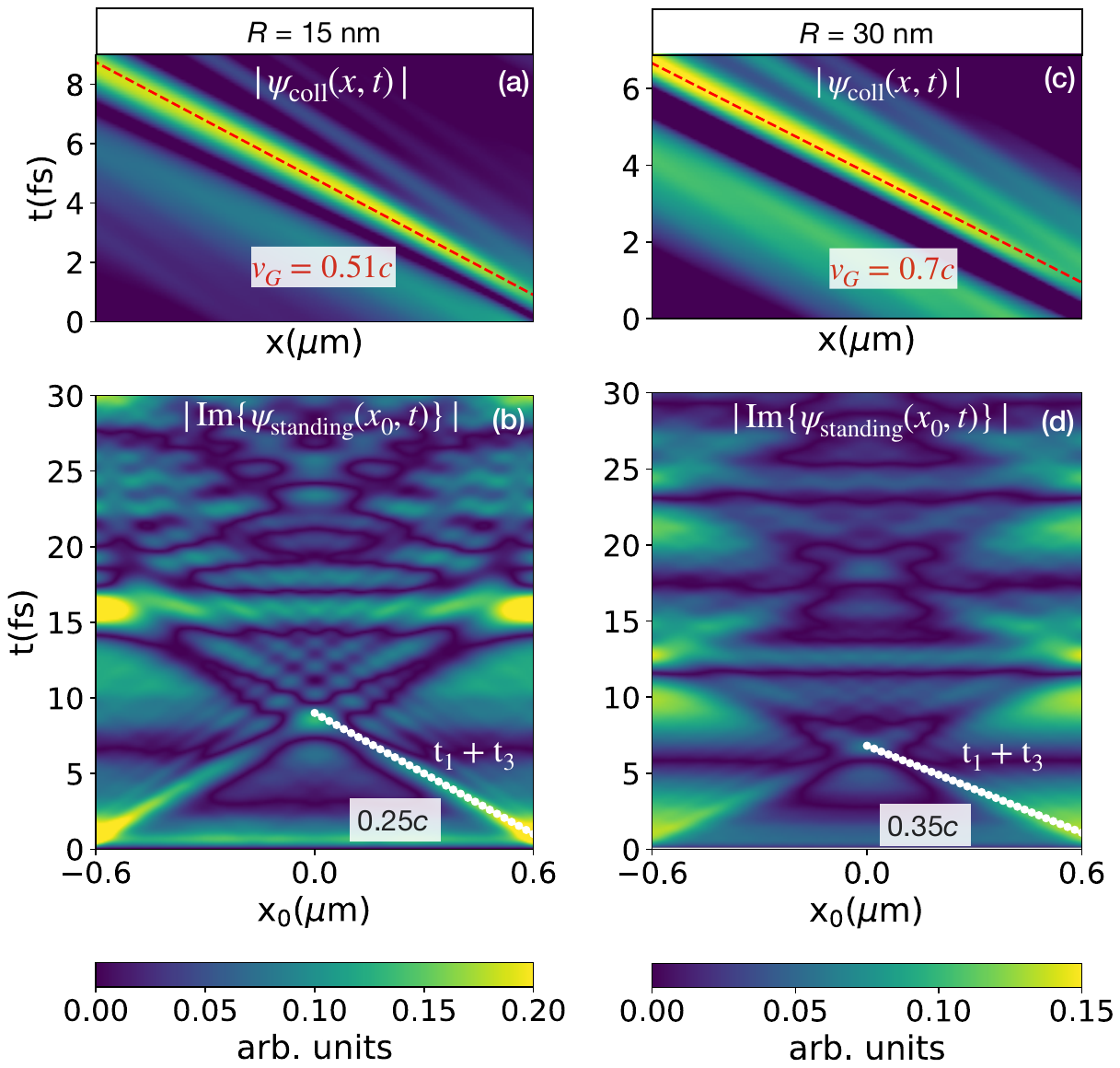}
    \caption{\textbf{Propagating modes and standing-wave model.} 
    \textbf{a,c} Superposition of propagating waves at the caps according to Eq.~\ref{eq:propagatingModes} for different modes together with the group velocity $v_G$ of the first SPP mode from DGTD computation.
    \textbf{b,d} Superposition of standing waves computed from Eq.~\ref{eq:standingWave} for different modes, together with the SPP round-trip time $t_1+t_3$. While panels a and b display the results for a wire with radius $R=15$~nm, panels c and d display results for a wire with radius $R=30$~nm.
    }
\label{fig:Figure5}
\end{figure}

We now proceed to a more detailed discussion of the SPP dynamics within a simplified toy model. By analyzing the dispersion relation of the SPPs in the finite wire (for the $n=0$ case, see the cross symbols in Fig.~\ref{fig:Figure2}d), we can simplify the map with reasonably good agreement by considering a sequence of $\omega_j$ and $k_j$ for each $j=1,\, 2,\, ...$ mode, where the values of the wave vector along the wire are determined by the standing-wave condition $k_j = j\pi/L$ for a wire of length $L$. In real space, such a combination of modes translates into a sum of plane waves of the form
\begin{equation} \label{eq:propagatingModes}
 \psi_{\rm coll}(x,t) = \sum_j A_j(x_0,y_0) \, \ee^{\ii(k_j x-\omega_j t)} + \text{c.c.},
\end{equation}
where $A_j(x_0,y_0)$ represents the weight of excited modes that depends on the trajectory $(x_0,y_0)$ of the exciting electron (see Fig.~\ref{fig:Figure2}e). To simplify the picture, we do not consider propagation losses (i.e., $\imag{\omega_j}=\imag{k_j}=0$), and take $A_j$ to be the amplitude of simulated EELS peaks. This facilitates a qualitative discussion of the collective behavior of the plasmonic modes inside the nanowire. In Fig.~\ref{fig:Figure5}a, we depict $\psi_{\rm coll}(x,t)$ for a set of $\{k_j,\omega_j\}$ evaluated from an electron trajectory at the right cap of the wire (i.e., at $x_0=L/2$ and $b=5$~nm), for the wire of radius $R=15$~nm studied in Fig.~\ref{fig:Figure2}f, and in Fig.~\ref{fig:Figure5}c for a $R=30$~nm wire. For the $R=15$~nm wire, we observe an initial feature originated at the right cap of the wire that propagates to the left cap at speed $\sim 0.5c$. This is the same that we already observed in Fig.~S6 in the SI, where the values of $|\vect{E}|$ along a line on the surface of the nanowire for the same excitation conditions are shown. Incidentally, in Figs.~\ref{fig:Figure5}a and ~\ref{fig:Figure5}c, we identify how satellites of the first plasmonic wavefront with smaller magnitude move at a slower speed to the left, where they eventually bounce at the caps (similar to those of Fig.~S6). These satellites are the result of the superposition of all modes $j$, which are excited by the electron. Altogether, the modeling provided by Eq.~\eqref{eq:propagatingModes} sheds light on how the different SPP modes propagate collectively as SPP wave packets in the wire and allows us to define a group velocity $v_G$.

Incidentally, for each mode, the EELS signatures in Figs.~\ref{fig:Figure2}b and ~\ref{fig:Figure2}f exhibit the shape of standing waves of the form $\cos^2{(k_j x)}$ located at $\omega = \omega_j$ and confined to the wire of length $L$. This is even more evident in Fig.~\ref{fig:Figure3}c, where we overlay the intensities of different modes from the experimental EEL spectra (solid-colored curves) with those from the toy model (dashed curves). To model such a behavior, we use
\begin{equation} \label{eq:standingWave}
 \psi_{\rm standing}(x,t) = \sum_j \hat{A}_j \cos^2(k_jx) \ee^{- \ii\omega_j t},
\end{equation}
here the amplitude $\hat{A}_j$ of the standing wave for mode $j$ is given by the maximum of $\Gamma_\EELS$ along the $x$-axis for each $j$ (see Fig.~\ref{fig:Figure3}). We display the corresponding results in Fig.~\ref{fig:Figure5}b that agree well with the simulation of $\tilde\Gamma_\EELS(t)$ shown in Fig.~\ref{fig:Figure2}h. This suggests that the SPP modes of order $n=0$ are responsible for the main features in the excitation probability maps for the $R=15$~nm wire. Similarly, we show in Fig.~\ref{fig:Figure5}d corresponding results for the $R=30$~nm wire and they, too, agree well with the experimentally observed features in Fig.~\ref{fig:Figure2}d.
Interestingly, such a simple toy model is related to the local density of optical states (LDOS) $\rho (\omega,x)=\sum_j \hat{A}_j \cos^2(k_jx) \,  \delta(\omega-\omega_j)$, where EELS and LDOS have been shown to share some common features \cite{Abajo_LDOS,hohenester2009electron,LK15}. Within this description, Fig.~\ref{fig:Figure5}b represents the Fourier transform in the time domain of the LDOS [i.e., $\rho (t,x)=\sum_j \hat{A}_j \cos^2(k_jx) \ee^{-\ii \omega t}$] and the result is therefore comparable to the excitation probability map $\tilde\Gamma_\EELS(t)$ in the time domain of Figs.~\ref{fig:Figure2}d and ~\ref{fig:Figure2}h. In fact, by considering the time $t_1+t_3$ of Figs.~\ref{fig:Figure4}f and ~\ref{fig:Figure4}g in Fig.~\ref{fig:Figure5}b, we observe that this time directly correlates with the main feature, thus relating $\rho (t,x)$ with the round-trip time $t_1+t_3$. 

Therefore, the description of the SPP modes with well-defined $\{k_j,\omega_j\}$ reproduces the main features of the $\tilde\Gamma_\EELS(t)$ map, thus characterizing both, the group velocity and the return time $t_R$. This result can be generalized for any long wire: for known wire length $L$, the values of $k_j=j\pi/L$ are also known, which in turn are related to $\omega_j$ through the dispersion relation of an infinitely long wire. Thereby, $v_G$ and $t_R$ are fixed with the prescription introduced above.

\subsection{Azimuthal modes and bulk contribution}
\begin{figure*}[t]
    \centering
    \includegraphics[trim = 5mm 79mm 5mm 48mm, clip, width=1
    \textwidth]{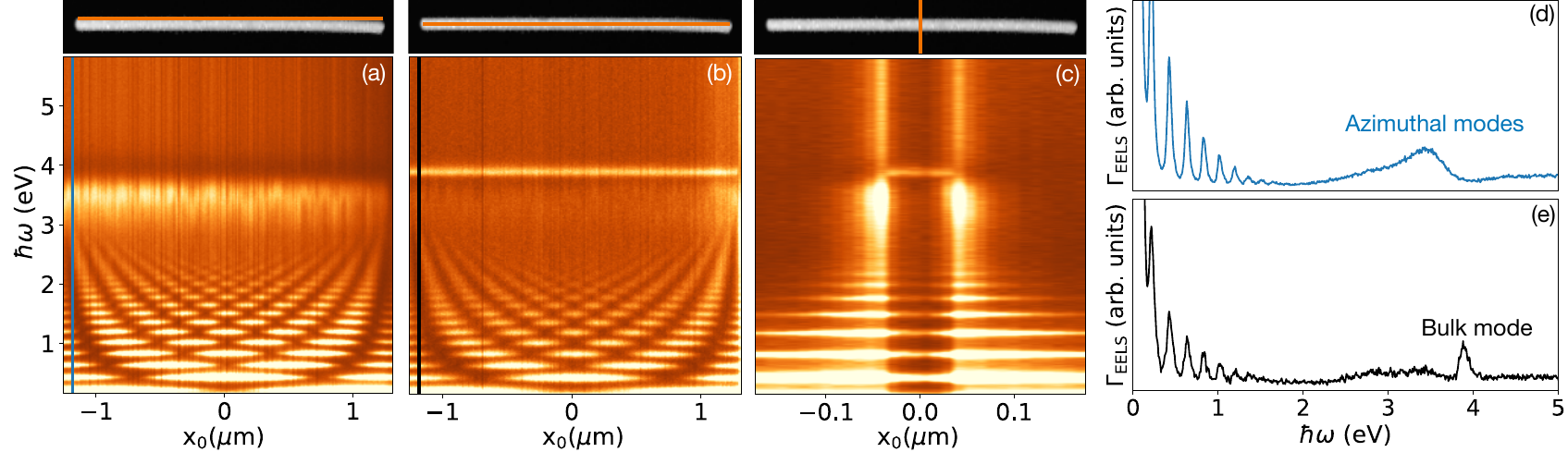}
    \caption{\textbf{The role of azimuthal and bulk modes.} 
    EELS of a nanowire (radius $R=30$~nm and length $L=2.6$~$\um$) excited by electrons with kinetic energy 200~keV. The EEL spectra are acquired for positions along the orange lines shown in the top insets.
    \textbf{a} EELS line scan for aloof electron trajectories parallel to the nanowire. 
    \textbf{b} EELS line scan for penetrating trajectories along the nanowire.
    \textbf{c} EELS line scan for trajectories vertically across the middle of the nanowire.
    \textbf{d} EEL spectrum for the position denoted with a blue line in (a), demonstrating the presence of higher-order azimuthal SPP modes.
    \textbf{e} EEL spectrum from a position denoted with the black line in (b), demonstrating the position of the bulk plasmon mode.
    }
\label{fig:Figure6}
\end{figure*}

So far, we have focused on the SPP dispersion relation of the fundamental azimuthal mode with $n=0$. Higher-order azimuthal SPP modes with $n \ge 1$ feature a dependence $\ee^{\ii n \varphi}$ for the associated electric fields and lie at higher energies. Thus far, little attention has been devoted to such higher-order modes and, to the best of our knowledge, only Refs.~\cite{zhang2011chiral, ChiralSPP2011} reported the experimental observation of azimuthal SPPs by quantum dot fluorescence imaging of silver nanowires with radii $\sim 150$~nm. Interestingly, we observe a very good agreement between the analytical dispersion relation (pink line) and the simulated one for the first-order azimuthal modes as displayed in Fig.~\ref{fig:Figure2}g. 


Furthermore, in Fig.~\ref{fig:Figure2}h, we observe a set of horizontal lines absent in the superposition of standing waves from the toy model in Fig.~\ref{fig:Figure5}b. This indicates the presence of azimuthal modes in the simulated results. This is further corroborated by the fact that these lines appear independent of the position of the electron path along the wire, thereby indicating that the corresponding modes do not propagate along but circle around the wire. To study the effect of azimuthal SPP modes in the $\tilde\Gamma_\EELS(t)$ map, we disentangle the individual contributions of the different modes to the map in Fig.~\ref{fig:Figure2}h. Specifically, as shown in Fig.~S1 of the SI, we compute the $\tilde\Gamma_\EELS(t)$ map by applying different energy filters to the original EEL spectra of Fig.~\ref{fig:Figure2}f. Interestingly, when including the first horizontal line of Fig.~\ref{fig:Figure2}f just above $\omega/\wp\sim 0.6$,  we retrieve the same features as in the experimental  $\tilde\Gamma_\EELS(t)$ map. When including the entire map (i.e., including the second horizontal line above $\omega/\wp\sim 0.6$ and below the surface plasmon energy $\omega/\wp = 1/\sqrt{2}$), the resulting features are those of a set of azimuthal waves that are closely spaced in energy and therefore yield interference and beating effects. The resulting band was observed in both experimental and simulated $\tilde\Gamma_\EELS(t)$ maps. However, a detailed experimental retrieval of higher-order azimuthal modes cannot be accomplished due to the presence of imperfections on the sample surface and the substrate that broaden the individual lines (see Fig.~\ref{fig:Figure6}a) \cite{note1}. 
Furthermore, in Sec.~S7 of the SI, we address the effect of the substrate on the EELS measurements, where the numerical calculations show indications of excited guided modes in the substrate. Additionally, in Fig.~S4 in the SI, we depict the fields on the surface of a finite wire at two particular instances of time, where the azimuthal SPP modes $n>0$ combine with fundamental SPP modes with $n=0$, thereby creating tilted wavefronts of the electric field. This indicates the presence of azimuthal modes with a nontrivial azimuthal dynamic around the wire in addition to the fundamental modes that only travel along the nanowire. 

When comparing the experimental EELS for the different sets of trajectories shown in Fig.~\ref{fig:Figure6}a-c, it is possible to directly compare the relative contribution of the so-called bulk plasmon \cite{bosman2007mapping, nelayah2007mapping, paper149} and the higher-order azimuthal SPP modes. For penetrating trajectories, we observe the bulk plasmon, which is clearly visible in Fig.~\ref{fig:Figure6}b and Fig.~\ref{fig:Figure6}c at energy $\hbar \omega \approx 3.8$~eV. As a matter of fact, as the bulk plasmon 
is a predominantly electronic excitation, its efficient excitation requires electron trajectories that penetrate the silver nanowire. In contrast, the excitation of the fundamental and higher-order azimuthal SPPs can be done efficiently via both penetrating and aloof electron trajectories, which are energetically located below the bulk plasmon. In the case of our thin wires, the higher-order SPP modes form an energy band around $\hbar \omega \sim 3.5$~eV and, therefore, are energetically rather close to the bulk plasmon mode. Nonetheless, as demonstrated in Figs.~\ref{fig:Figure6}d and ~\ref{fig:Figure6}e, we are able to distinguish between higher-order azimuthal SPP modes and the bulk plasmon mode when comparing the EELS signals from aloof and penetrating positions. Specifically, in the EEL spectrum for the aloof trajectory in Fig.~\ref{fig:Figure6}d, we observe a relatively wide peak around $\hbar \omega \sim 3.5$~eV related to the higher-order azimuthal SPP modes and no signal at the bulk plasmon energy. In contrast, in the EEL spectrum for the penetrating trajectory in Fig.~\ref{fig:Figure6}e, we observe a rather sharp peak at the bulk plasmon energy $\hbar \omega \sim 3.8$~eV.
Similarly, in Fig.~S3c in the SI, we observe in the simulated EEL spectra the presence of the bulk plasmon peak only for the penetrating trajectory and it is located exactly at the plasmon energy of the Drude model (i.e., $\hbar \omega_{\rm p} \sim 9.17$~eV). Note that there is a discrepancy between the experimental bulk plasmon peak and simulated one, which could be improved by using measured optical data in the numerical computations, where we neglected the screening effects of interband transitions.


\section{Conclusion}
In summary, we have introduced a combined experimental-theoretical approach for a time-dependent perspective on EELS signals from plasmonic nanostructures. This time-dependent perspective allows for identifying and unraveling the dynamics of different plasmonic excitations and leads to deeper insights into the underlying physics. 
In particular, our study highlighted the role of propagating fundamental surface plasmon polariton wave packets in thin silver nanowires that move at specific group velocities. We have been able to quantitatively determine the group velocity of these wave packets based only on the radius and material properties of the wire. The SPP wave packets bounce back and forth in the wire and lead to an enhancement of light-matter interaction. Moreover, and despite the fact that the thin wire represents an almost one-dimensional system, we have identified the effects of higher-order azimuthal surface-plasmon polaritons in both, DGTD-simulated and measured EEL data. 
%
This has been accomplished by going beyond the study of the spatio-spectral information obtained via standard EELS maps. In fact, the Fourier transform of the EELS signal sheds light on the electron-photon interaction probability mechanisms in the time domain, and this approach can be applied to any material configuration. For instance, such a comprehensive understanding of the temporal evolution of plasmonic excitations opens a route for controlling the losses in plasmonic systems. Specifically, while the overall result of the interaction of the electron with the plasmonic nanostructure is an energy loss, the oscillatory character of the electromagnetic fields provides a local/differential gain that can be exploited via more complex configurations than we have considered in this paper, i.e. by properly "surfing" on the SPP dynamics. In doing so, our time-dependent perspective provides a refined view of losses in metals that can be used to design better plasmonic resonators \cite{Gonçalves_2020} for chemical applications, plasmonic sensors for spectroscopy, and controlling coherent effects \cite{talebi2017interaction}.

\section{Methods}

\subsection{Sample preparation} \label{Sec:experiments_methods}


Materials: Silver nitrate (99\%), ethylene glycol (99.8\%), poly-(vinylpyrrolidone) (Mw = 40000), tetraethyl orthosilicate (TEOS),sodium chloride, ammonia solution (28 wt \%), and absolute ethanol were purchased from Aldrich. All chemicals were used without further purification.
 
Silver nanowires (AgNWs) of approximately 1--3~µm length and 30 nm width were synthesized according to our previous report \cite{sample2021}, which was adapted from a method developed by Hu \textit{et al.} \cite{Hu2012RapidCH}. In short, 34 mL poly-(vinylpyrrolidone) solution (1.47 mM in ethylene glycol) was heated to 160°C, then 40 µL sodium chloride solution (also dissolved in ethylene glycol, 0.1 M) was added. After stirring for 1~min, 6 mL silver nitrate solution (in ethylene glycol, 0.67 M) was added at a rate of 10 µL/s until the mixture appeared turbid. Once turbid, the rest of the silver nitrate solution was added rapidly. The mixture was stirred at 160°C for 30~mins, then cooled to room temperature and centrifuged. The supernatant was discarded and the silver nanowires were redispersed in ethanol. The dispersion was stored at 4°C until use. Samples of Ag nanowires were prepared for characterisation by depositing a drop of the dispersion on silicon nitride supported TEM grids of 15nm thickness (purchased from Plano) and letting them dry at room temperature.

\subsection{Electron microscopy} \label{Sec:microscopyMethods}
STEM imaging and electron energy-loss spectroscopy were performed on the Nion aberration-corrected high energy resolution monochromated EELS-STEM (HERMES) system \cite{Krivanek_2014} at Humboldt-Universität zu Berlin operated at 30~kV, 60~kV, and 200~kV accelerating voltage, equipped with a Dectris ELA direct detector. The energy resolution of the EELS acquisition was measured using FWHM of the zero-loss peak (ZLP) and was found to be between 12--21~meV depending on accelerating voltage.  The convergence semi-angle was 10 mrad. EELS was performed using a 5--20~nm pixel size with a maximum field of view of $5$~$\um$.
All analysis of the microscopy data was carried out using the open-source Python package HyperSpy \cite{Pea2021hyperspyhyperspyRV}. Richardson-Lucy deconvolution was employed for ZLP removal within the HyperSpy environment.

\subsection{Numerical simulation}  \label{Sec:DGTD_methods}


The DGTD finite-element approach \cite{DGTD} was used to simulate the electromagnetic problem of a silver nanowire excited by an electron beam. In this approach, a polynomial spatial interpolation basis of adjustable order (typically third or fourth order) on an unstructured tetrahedral mesh (see Fig.~S7 in the SI) is combined with a low-storage Runge-Kutta time integrator (typically fourth order), thereby providing an efficient solver for the time-dependent Maxwell's equations 
\begin{subequations} \label{Eq:Maxwell}
\begin{align}     
    \partial_t \vect{H} (\vect{r}, t) = &   
    - \mu^{-1}_0\mu^{-1}(\vect{r}) \,  \vecnabla \times \vect{E} (\vect{r}, t),
    \\
    \partial_t \vect{E} (\vect{r}, t)=&  
    \, \varepsilon^{-1}_0\varepsilon^{-1}(\vect{r}) \, \left[ 
    \vecnabla \times\vect{H} (\vect{r}, t) - \vect{j}(\vect{r}, t)\right].
\end{align}
\end{subequations}
Here $\vect{j}$ is the total current density that encompasses any currents associated with the excitation source and dispersive polarization currents (see Sec.~S6 in the SI for further technical details). The resulting method is memory-efficient and particularly well-suited for calculating broadband spectra.

Due to the symmetry of the nanowire, we only have to consider trajectories along a line on the right half side of the nanowire with respect to the $yz$-plane, which cuts through the middle of the nanowire (see Fig.~\ref{fig:Figure2}e). At the same time, we have cylindrical symmetry regarding the rotation along the long axis of the finite cylinder, which means the rotation of the trajectories along the long axis will give the same EELS maps. \cite{paper149}.

{\bf Electron source.} In particular, the electron source is implemented as a Gaussian charge distribution of the form
\begin{equation}   \label{eq:Gaussiancharge}
    \rho(\vect{r}) = - \frac{ e}{\sigma_e^3 \sqrt{\pi^3} } \ee ^{-r^2/\sigma_e^2},
\end{equation}
with width $\sigma_e=5$\,nm. This choice essentially prevents numerical artifacts arising from the implementation of a point-charge particle moving inside the nanowire. The induced electric fields $\Eb^\ind$ along the electron trajectory are recorded in the time domain, and subsequently, an on-the-fly Fourier transform within DGTD allows the conversion of the simulation data to the frequency domain. EEL spectra for the nanowire are then computed in a post-processing step according to Eq.~\eqref{eq:EELP_dw}.

{\bf Non-recoil approximation.} By integrating $\Gamma_\EELS (\omega)$ following Eq.~\eqref{eq:energyLoss}, we calculated the total energy loss $\Delta E$ of the electron to the various plasmonic excitations, which is much smaller than electrons' kinetic energy. Additionally, the total momentum transfer from the electron to the nanowire is $\hbar k \approx \hbar \omega_{\rm loss}/v \ll m_ev$, where $m_e$ is the mass of the electron. Therefore, the no-recoil approximation \cite{paper149} is adopted for the numerical computations.
\section*{Acknowledgments}
%
A.~E., T.~K., K.~B., and C.~T.~K. acknowledge funding by the German Research Foundation (DFG) in the framework of the Collaborative Research Center (CRC) 951 (Project ID 182087777).
%
%
H.~C.~N. acknowledges funding by the DFG, Project No.449639588 (NE 2491/2-1) in the framework of the DFG Priority Programme 2244: 2D Materials – Physics of van der Waals [hetero]structures. 
%
%
K.~B. acknowledges funding by the German Research Foundation (DFG) in the framework of the CRC 1636 (Project ID 510943930 - Project A06) and the CRC 1375 (Project ID 398816777 - Project A06). 
C.~T.~K. acknowledges funding by DFG in Projekt CAPTURE (Project ID 541930824).
\vspace{0.4cm}
\section*{Author contribution}
A.~E. initiated the study. K.~B. and C.~T.~K. supervised the research.
W.~Z. carried out the DGTD computations and contributed to the analysis of the experimental results and the theoretical analysis. 
A.~R.~E. contributed to the theoretical analysis and coordinated the collaboration between the theoretical and the experimental teams. 
A.~E. carried out the experiments and experimental EELS data processing. 
B.~H. supported in the operation of the microscope. 
H.~C.~N. contributed to the experimental data processing. 
T.~K. helped with the DGTD computations and advised on the theoretical analysis. 
H.~H and Y.~L. provided the Ag nanowire samples.
All authors participated in the discussion of the results and the writing of the manuscript.
\hfill

%

\end{document}


\title{Real-time surface plasmon polariton propagation in silver nanowires  \\ {\color{gray} \small -- SUPPLEMENTARY INFORMATION --}}

%
%
%
%
%
%

\author{Wenhua~Zhao\,\orcidlink{0009-0004-5721-607X}}
\affiliation{Max-Born-Institut, 12489 Berlin, Germany}
\affiliation{Department of Physics, AG Theoretische Optik \& Photonik, Humboldt-Universität zu Berlin, 12489 Berlin, Germany}
\author{Álvaro~Rodríguez~Echarri\,\orcidlink{0000-0003-4634-985X}}
\affiliation{Max-Born-Institut, 12489 Berlin, Germany}
%
\author{Alberto~Eljarrat\,\orcidlink{0000-0002-0968-5195}}
\affiliation{Department of Physics, Structure Research and Electron Microscopy group, Humboldt-Universität zu Berlin, 12489 Berlin, Germany}
\affiliation{Center for the Science of Materials Berlin, Humboldt-Universität zu Berlin, 12489 Berlin, Germany}
\author{Hannah~C.~Nerl\,\orcidlink{0000-0003-4814-7362}}
\affiliation{Department of Physics, Structure Research and Electron Microscopy group, Humboldt-Universität zu Berlin, 12489 Berlin, Germany}
\author{Thomas~Kiel\,\orcidlink{0000-0002-6070-9359}}
\affiliation{Department of Physics, AG Theoretische Optik \& Photonik, Humboldt-Universität zu Berlin, 12489 Berlin, Germany}
\author{Benedikt~Haas\,\orcidlink{0000-0002-9301-8511}}
\affiliation{Department of Physics, Structure Research and Electron Microscopy group, Humboldt-Universität zu Berlin, 12489 Berlin, Germany}
\affiliation{Center for the Science of Materials Berlin, Humboldt-Universität zu Berlin, 12489 Berlin, Germany}
\author{Henry~Halim\,}
\affiliation{Helmholtz-Zentrum Berlin for Materials and Energy,
	14109 Berlin, Germany}
\author{Yan~Lu\,}
\affiliation{Helmholtz-Zentrum Berlin for Materials and Energy,
	14109 Berlin, Germany}
\affiliation{Friedrich-Schiller-University, 07737 Jena, Germany}
\author{Kurt~Busch\,\orcidlink{0000-0003-0076-8522}}
\email{Corresponding author: kurt.busch@physik.hu-berlin.de}
\affiliation{Department of Physics, AG Theoretische Optik \& Photonik, Humboldt-Universität zu Berlin, 12489 Berlin, Germany}
\affiliation{Max-Born-Institut, 12489 Berlin, Germany}
\author{Christoph~T.~Koch\,\orcidlink{0000-0002-3984-1523}}
\email{Corresponding author: Christoph.Koch@hu-berlin.de}
\affiliation{Department of Physics, Structure Research and Electron Microscopy group, Humboldt-Universität zu Berlin, 12489 Berlin, Germany}
\affiliation{Center for the Science of Materials Berlin, Humboldt-Universität zu Berlin, 12489 Berlin, Germany}
\textbf{}

\begin{abstract}
    \bf In this document, we provide technical details and additional information regarding our study entitled "Real-time surface plasmon polariton propagation in silver nanowires". Specifically, we give a derivation of the temporal analysis of EELS, address the dispersion relation of infinite nanowires, expand the technical formalism to solve the electromagnetic problem numerically in the DGTD formalism, and write further details about the characteristics of the SSPs modes in silver long nanowires. Finally, the influence of the substrate is addressed.
\end{abstract}

\date{\today}
\maketitle
\tableofcontents

\newpage

\section{Fourier transform of EELS}
First, we calculate the time-domain evolution of the electron energy loss probability
\begin{equation}
    \Gamma_\mathrm{EELS} (\omega) = \frac{e}{\pi\hbar\omega} \int dz\, \real{ \ee^{-\ii\omega \frac{z-z_0}{v}} E_{z}^\ind (z, \omega) }, \quad \quad \omega \ge 0,
\end{equation}
where the integral runs over the electron trajectory.
By means of the inverse Fourier transform, we obtain
\begin{equation}
    \tilde\Gamma_\EELS(t) = \int_{-\infty}^{\infty} \frac{d\omega}{2\pi}\, \Gamma_\EELS(\omega)  \theta(\omega)\,\ee^{-\ii \omega t} = \frac{e}{\hbar} \int_{-\infty}^{\infty} \frac{d\omega}{2\pi}\,  \frac{1}{\pi \omega} \int_{-\infty}^{\infty} dz \; \real{ E(z,\omega) \ee^{-\ii \omega \frac{z-z_0}{v}}} \, \ee^{-\ii \omega t} \theta(\omega),
\end{equation}
where we have $E_{z}^\ind \to E$ in order to simplify the notation. In addition, we have introduced the Heaviside step function $\theta(\omega)$ to guarantee that $ \Gamma_\mathrm{EELS} (\omega)$ is defined for $\omega \ge 0$. After rearranging terms, and using the fact that the electric field in the time domain is a real-valued [i.e., implying $E^*(z,-\omega)=E(z,\omega)$], 
we find
\begin{equation}
    \tilde\Gamma_\EELS(t) = \int_{0}^{\infty} \frac{d\omega}{2\pi}\,  \frac{e}{\hbar\omega\pi} \int_{-\infty}^{\infty} dz \; \real{ E(z,\omega) \ee^{-\ii \omega t'}} \, \cos(\omega t)-
    \frac{\ii}{2} \int_{-\infty}^{\infty} \frac{d\omega}{2\pi}\,  \frac{e}{\hbar\omega\pi} \int_{-\infty}^{\infty} dz \; E(z,\omega) \ee^{-\ii \omega t'} \, \sin(\omega t).
\end{equation}
In this expression, we have defined $t'=(z-z_0)/v$ so that $t'$ represents the arrival time of the electron at position $z$ when starting at $z_0$ at moving with constant velocity $v$. Now, we focus on the imaginary part and observe that the frequency integration over can be understood as a Fourier transform, i.e. as $\mathrm{FT} \; \{ E(z,\omega) \cdot \sin(\omega t) / \omega \}$ to the $t'$-domain $\omega \rightarrow t' =(z-z_0)/v$. With the help of the convolution theorem, we obtain 
\begin{align}
     \imag{\tilde\Gamma_\EELS(t)} &= -\frac{e}{2\pi \hbar}\int_{-\infty}^{\infty} dz \, E(z, \tilde t) * \mathrm{FT} \; \left\{\frac{\mathrm{sin}(\omega t)}{\omega} \right\} (\tilde t) \bigg|_{\tilde t = t' }.
\end{align}
Further, we assert that $\mathrm{FT} \; \{\sin(\omega t)/\omega\} = {\rm rec}\left({\tilde t}/2t\right)/2$, where ${\rm rec}(x/W)$ denotes the rectangular window function of width $W$. As a result, we obtain 
\begin{align}
     \imag{\tilde\Gamma_\EELS(t)} &= -\frac{e}{4 \pi \hbar}\int_{-\infty}^{\infty} dz \, E(z, \tilde t) * {\rm rec}\left(\frac{\tilde t}{2t}\right) \bigg|_{\tilde t = t' } \\
  & = -\frac{e}{4 \pi \hbar}\int_{-\infty}^{\infty} dz \; \int_{t' -t}^{t' +t} d\tau \; E(z, \tau).
\end{align}
The above result provides an alternative interpretation:
Instead of integrating the induced electric field $E$ at the position of the electron as for $\Gamma_\EELS(\omega)$, we integrate, for each position, over a time window around the arrival time $t' $ with a width of $2t$. 

\begin{figure}
\includegraphics[trim = 0mm 28mm 0mm 0mm, clip, width=1
    \textwidth]{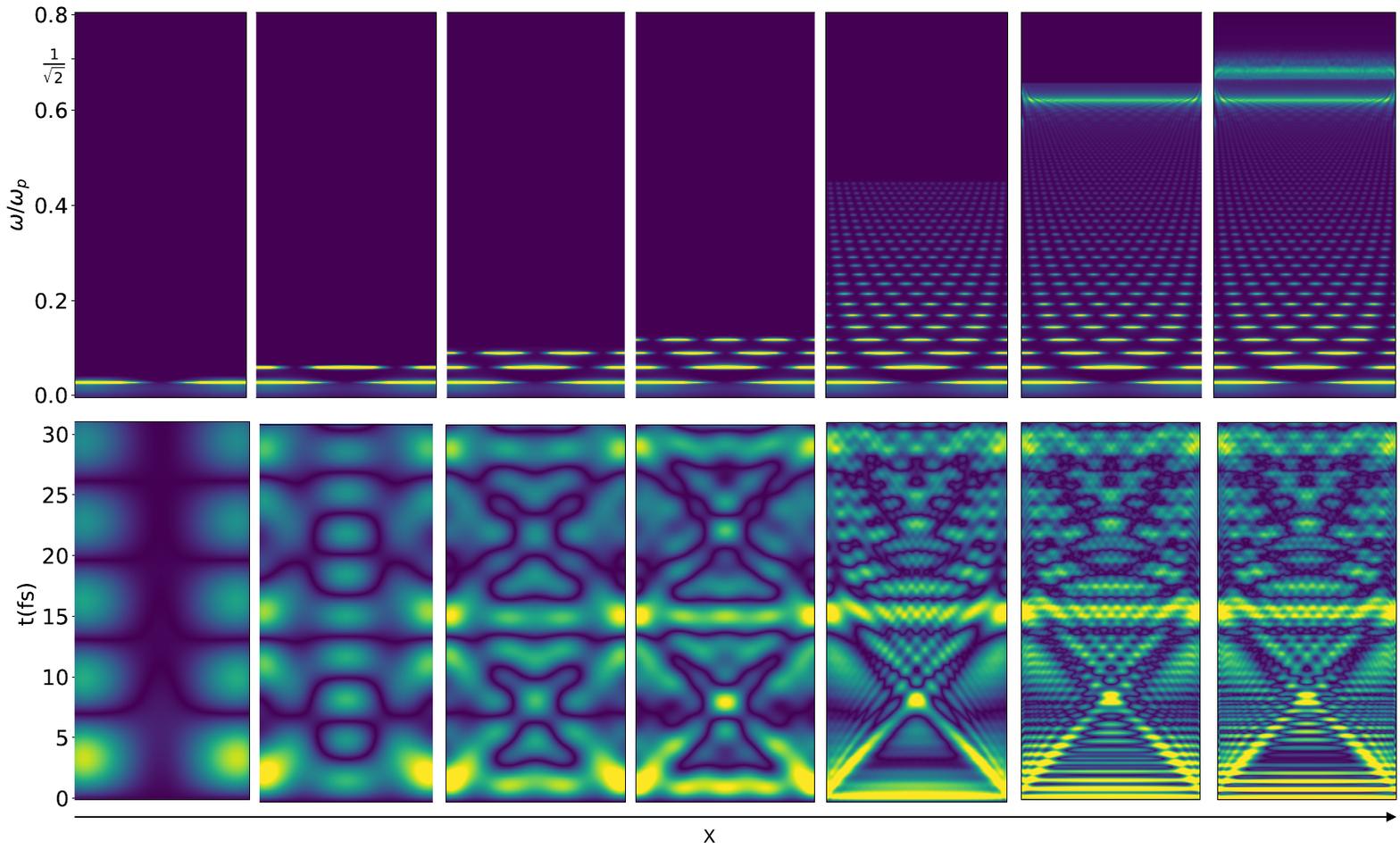}
    \caption{\textbf{Filtered Fourier transforms of the simulated EEL spectra:} 
    The full EELS simulations for a silver nanowire (radius $R=15$~nm and length $L=1.2$~$\um$) are depicted in Fig.~2f of the main text (see caption for further details). Here, different regions of the EELS map are masked so that the Fourier transformation of the different filtered EELS maps into the time domain help to identify the role of the different parts of the EEL spectra. Upper row)  EELS map by filtering different modes. Lower row) Amplitude of the imaginary part of the FT $|\imag{\tilde\Gamma_\EELS(t)}|$ for the corresponding filtered EELS map displayed on the upper row.
    }
    \label{fig:S1}
\end{figure}

\begin{figure}
\includegraphics[trim = 20mm 63.5mm 18mm 63mm, clip, width=1
    \textwidth]{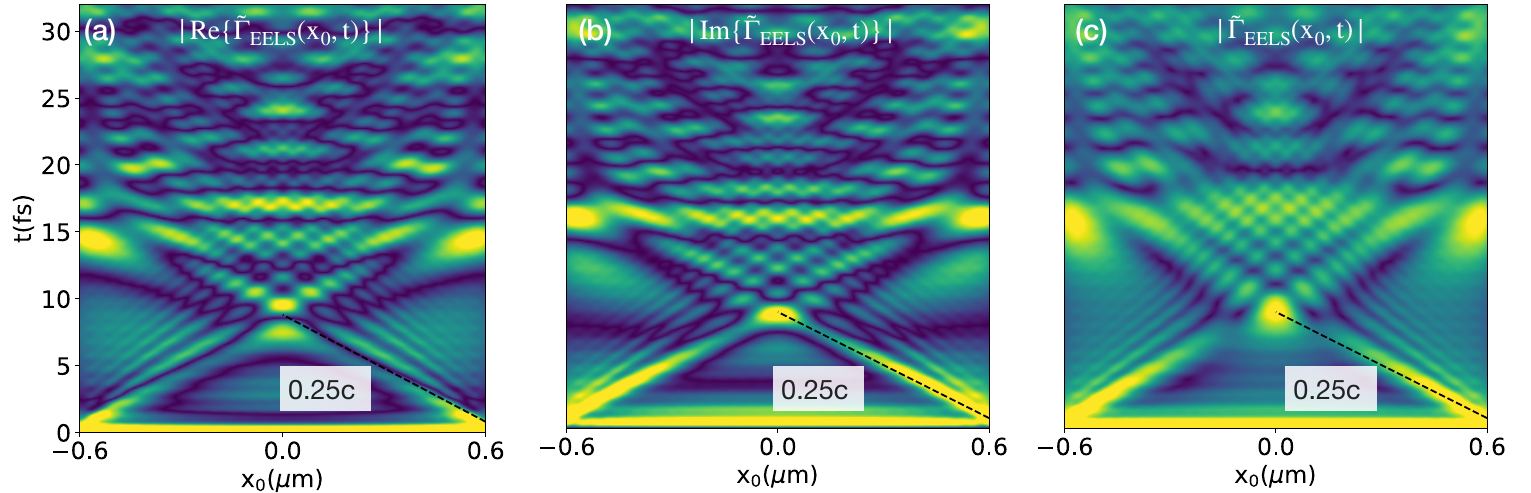}
    \caption{\textbf{Excitation probability maps}. For the EELS simulations displayed in Fig.~2f of the main text (silver nanowire, radius $R=15$~nm and length $L=1.2$~$\um$), we compute $\tilde\Gamma_\EELS(t)$ from a spectrum without azimuthal SPP modes. In \textbf{a}, we display the real part, in \textbf{b} the imaginary part, and in \textbf{c} the absolute value. Using the results of the imaginary part, we obtain a propagation speed of $0.25c$ for the main feature (black dashed lines).
    }     
    \label{fig:S1_new}
\end{figure}

In the main text, we have discussed the resulting $\tilde\Gamma_\EELS(t)$ maps as a function of the electron trajectory position $x_0$ for different positions near the silver nanowire location. Interestingly, we can take $\Gamma_\EELS(\omega)$ and apply the Fourier transform filtering particular parts of the spectrum to identify the role of each part. We show in Fig.~\ref{fig:S1} the computed $|\imag{\tilde\Gamma_\EELS(t)}|$ by applying a frequency filter to the original $\Gamma_\EELS(\omega)$. The first row of Fig.~\ref{fig:S1} shows the modes we considered (i.e., the dark blue color is discarded), and the second row of the figure shows the corresponding Fourier transform $|\imag{\tilde\Gamma_\EELS(t)}|$. We note that, by considering one mode, we obtain a standing wave in the Fourier-transformed signal, while adding more modes leads to a feature propagating in $|\imag{\tilde\Gamma_\EELS(t)}|$ along the nanowire. Interestingly, the inclusion of the azimuthal SPP modes yields horizontal lines in the Fourier transform. Moreover, we show in Fig.~\ref{fig:S1_new} the real and imaginary part of 
$\tilde\Gamma_\EELS(t)$.


\newpage
\section{Dispersion relation of surface plasmon polaritons in an infinite cylinder}
\begin{figure}[ht]
	\centering
	\includegraphics[trim = 0mm 73mm 0mm 40mm, clip, width=1
    \textwidth]{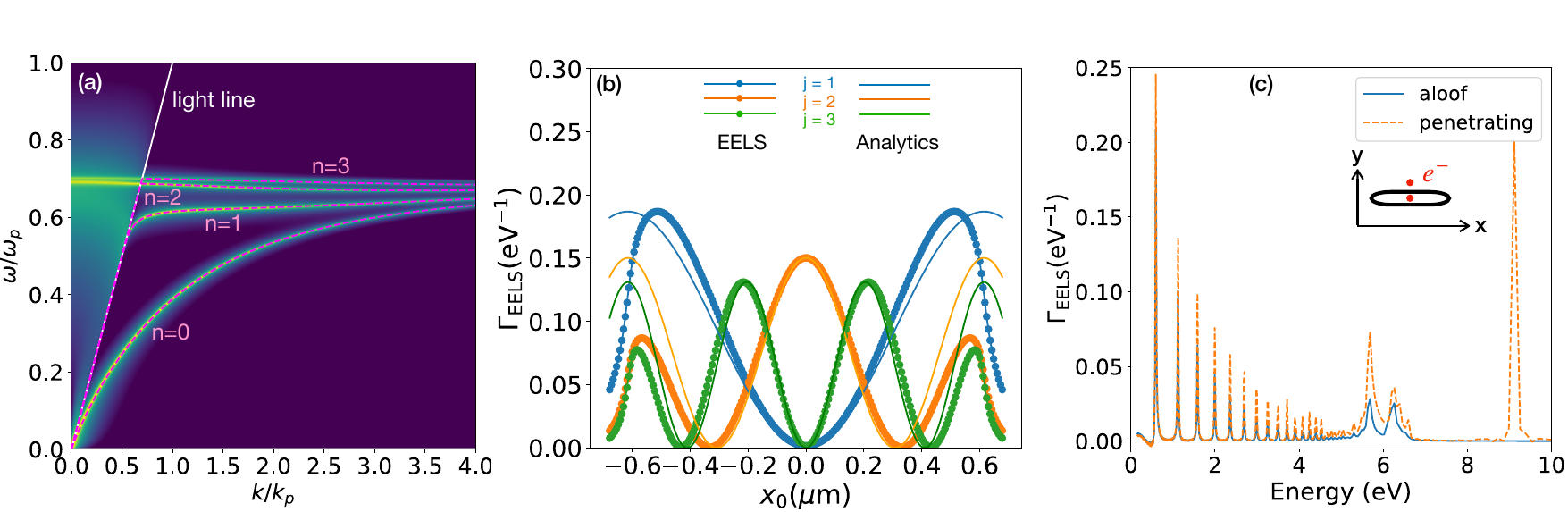}
	\caption{
    \textbf{a} Normalized EELS for an infinite cylinder of radius $R=15$~nm, made of silver by considering Drude model and $b=5$~nm \cite{talebi2015excitation}. Pink lines are calculated analytically from \cite{AE1974}, which correspond to different azimuthal orders $n=0,1,2$, and $3$.
    \textbf{b} The first 3 modes in the EELS map of Fig.~1f in the main text (dotted lines) compared with the analytical standing waves model $ \propto \cos^2(k_jx)$ with $k_j=j\pi/L$ (solid lines).
    \textbf{c} EELS for a nanowire of length $L=1.2$~$\um$ and radius $R=15$~nm, where the electron beam interrogates both an aloof trajectory and a penetrating one (see red dots in the sketch) located in the center of the nanowire. Note the presence of the bulk mode at energies around $\hbar\omega=\hbar\wp=9.17$~eV for the penetrating trajectory.
 }
\label{fig:S2}
\end{figure}

The analytical dispersion relation of surface plasmon 
polaritons for an infinitely long cylinder is given as (see 
Ref.~\cite{AE1974}).
\begin{equation} \label{eq:analyDispersion}
  \nu ^2 \nu'^{2}  (\nu' \varepsilon \alpha_n - \nu \varepsilon' \beta_n)(\nu' \alpha_n - \nu \beta_n) - \frac{n^2}{a^2} (\varepsilon' - \varepsilon)^2 k^2 \frac{\omega ^2}{c^2} = 0.
\end{equation}
Here, using the original notation of Ref.~\cite{AE1974}, we define
\begin{align}
    \nu &= (k^2 - \varepsilon \omega^2/c^2)^{\frac{1}{2}},  & \alpha_n &= \frac{\rm d}{{\rm d}(\nu a)} {\rm ln} \, I_n (\nu a),  \\ 
    \nu' &= (k^2 - \varepsilon \omega'^{2}/c^2)^{\frac{1}{2}}, & \beta_n &= \frac{\rm d}{{\rm d}(\nu' a)} {\rm ln} \, K_n(\nu' a) , 
\end{align}
where $I_n$ and $K_n$ denote the modified Bessel functions of the first and second kind of order $n$, respectively. Further, the angular frequency is $\omega$ and $k$ stands for the wave number along the direction of the wire. 
In Fig.~\ref{fig:S2}a, we show for $n=0,...,3$ the solution of Eq.~\eqref{eq:analyDispersion} for a silver wire with plasma
frequency (energy) $\hbar \wp = 9.17$~eV 
and radius $R=15$~nm (pink dashed lines), normalized to the corresponding
wave number $\kp=\wp/c$. The analytical solution of the modes is 
superimposed onto the EELS map computed for the same system and obtained 
from Ref.~\cite{talebi2015excitation} (arbitrary units) for an aloof
electron trajectory with an impact parameter $b=5$~nm (distance from 
the cylinder surface).

\newpage
\section{Azimuthal SPP modes and influence of the nanowire terminations}
\begin{figure}[h!]
\includegraphics[trim = 13mm 14mm 15mm 41mm, clip, width=0.95
    \textwidth]{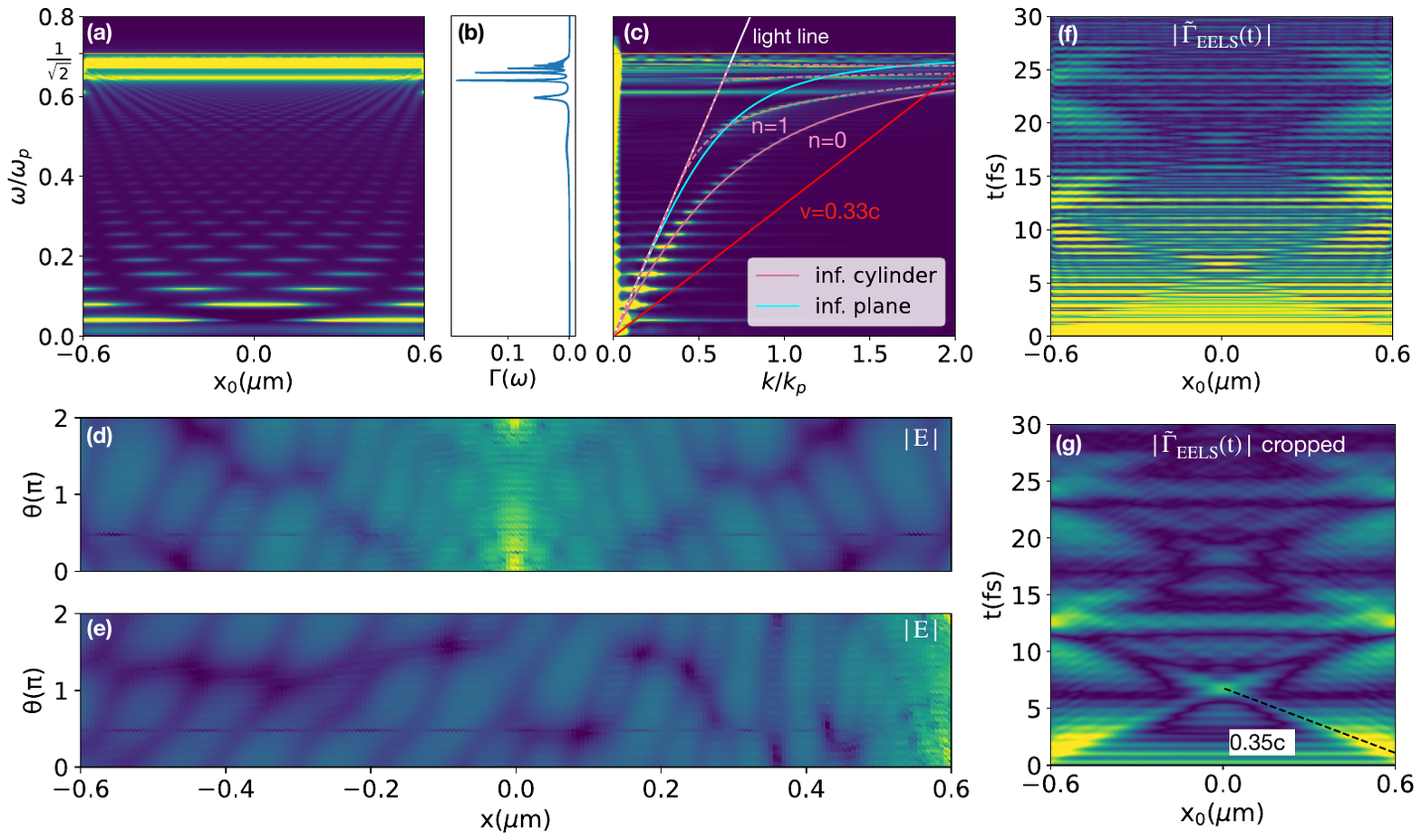}
    \caption{\textbf{Analysis of the azimuthal SPP modes.}
    \textbf{a} EELS map for a silver nanowire of length $L=1.2$~$\um$ and radius $R=30$~nm (similar as in Fig.~2e in the main text)
    \textbf{b} EELS of a silver sphere of radius $30$~nm excited with an 
    aloof electron beam with velocity $0.33c$ and impact parameter $5$~nm,
    using the MNPBEM toolbox \cite{MNPBEM}.
    \textbf{c} Dispersion relation through a spatial Fourier transform of 
    (a) along the $x$-axis.
    \textbf{d} Electric fields on the surface of the nanowire as a function 
    of position $x$ and azimuthal angle $\theta$, for a nanowire of radius $30\,$nm and length $1.2\rm \mu m$ using an aloof electron beam 
    (impact parameter $5\,$nm)
    passing the middle of the nanowire ($x_0=0\,$nm). 
    \textbf{e} Same as (d) but with an aloof electron beam traversing
    (impact parameter $5\,$nm) passing at the right end of the nanowire ($x_0=600\,$nm).
    \textbf{f} Absolute value of the temporal Fourier transform $\tilde\Gamma_\EELS(t)$  of the EELS map in (a).
    \textbf{g} Same as f, but cropping panel (c) to remove the influence 
    of the azimuthal SPP modes so that a cleaner image is obtained. This
    allows for an easier determination of the slope of the feature which
    moves towards the center with velocity $\sim 0.35c$ (black dashed line).
    }
    \label{fig:S3}
\end{figure}

As shown in Fig.~\ref{fig:S3}c, we obtain the dispersion relation of the SPPs by carrying out the Fourier transform of the EELS 
map (see Fig.~\ref{fig:S3}a) for a long silver nanowire (length $L=1.2$~$\um$ and radius $R=30$~nm).
However, we observe that there exist two bright horizontal lines around the frequency ($0.6<\omega/\wp<1/\sqrt{2}$) for the $n=1$ azimuthal SPP mode, 
which we do not expect for an infinitely long cylindrical nanowire (e.g., see  Fig.~\ref{fig:S2}). Interestingly, given that the long nanowire has hemispherical caps, we plot in Fig.~\ref{fig:S3}b the EEL spectrum 
of a silver sphere of radius $30\,$nm excited by an aloof electron beam 
(impact parameter $b=5$~nm) with velocity $v=0.33c$. 
Note, that in
the spatially resolved EELS map of Fig.~\ref{fig:S3}a, the positions 
wie find the Mie resonances of of Fig.~\ref{fig:S3}b are at the wire edges, 
i.e. where the hemispheres caps are located. In other words the spectral 
position of the sphere resonances (Fig.~\ref{fig:S3}b) are thus consistent
with the edge features of the hemispherically capped nanowire.


In addition, the azimuthal SPP modes of the nanowire, which exist on the wire surface and move along the azimuthal direction (the cylindrical circumference), can be observed in Fig.~\ref{fig:S3}d-e. There, we depict the electric field amplitudes (arb. units) on the surface of the nanowire as a function of the propagation direction $x$ and the azimuthal angle $\theta$, for a nanowire of radius $30\,$nm and length $1.2$~$\um$ with an electron beam traversing (d) in the center of the wire (i.e., $x_0=0$) and (e) at the right end of the nanowire (i.e., $x_0=L/2$) with an impact parameter of $b=5$~nm  (i.e., in both (d) and
(e), we have $y_0=35$~nm). The fields are recorded at a time after 
the electron has passed the nanowire. We clearly see that due to the 
azimuthal motion of the $n>0$ higher-order azimuthal modes, the electric field oscillation travels with a wave vector oblique to the long axis of the 
nanowire. Because the fundamental $n=0$ SPP modes travel strictly parallel 
to the cylinder axis, this oblique wave fronts are direct evidence of the
excitation of the higher-order $n>0$ azimuthal SPP modes.

\newpage
\section{Role of interband transitions and damping in the Drude model}
\begin{figure}[h!]
\includegraphics[trim = 0mm 28mm 0mm 35mm, clip, width=1
    \textwidth]{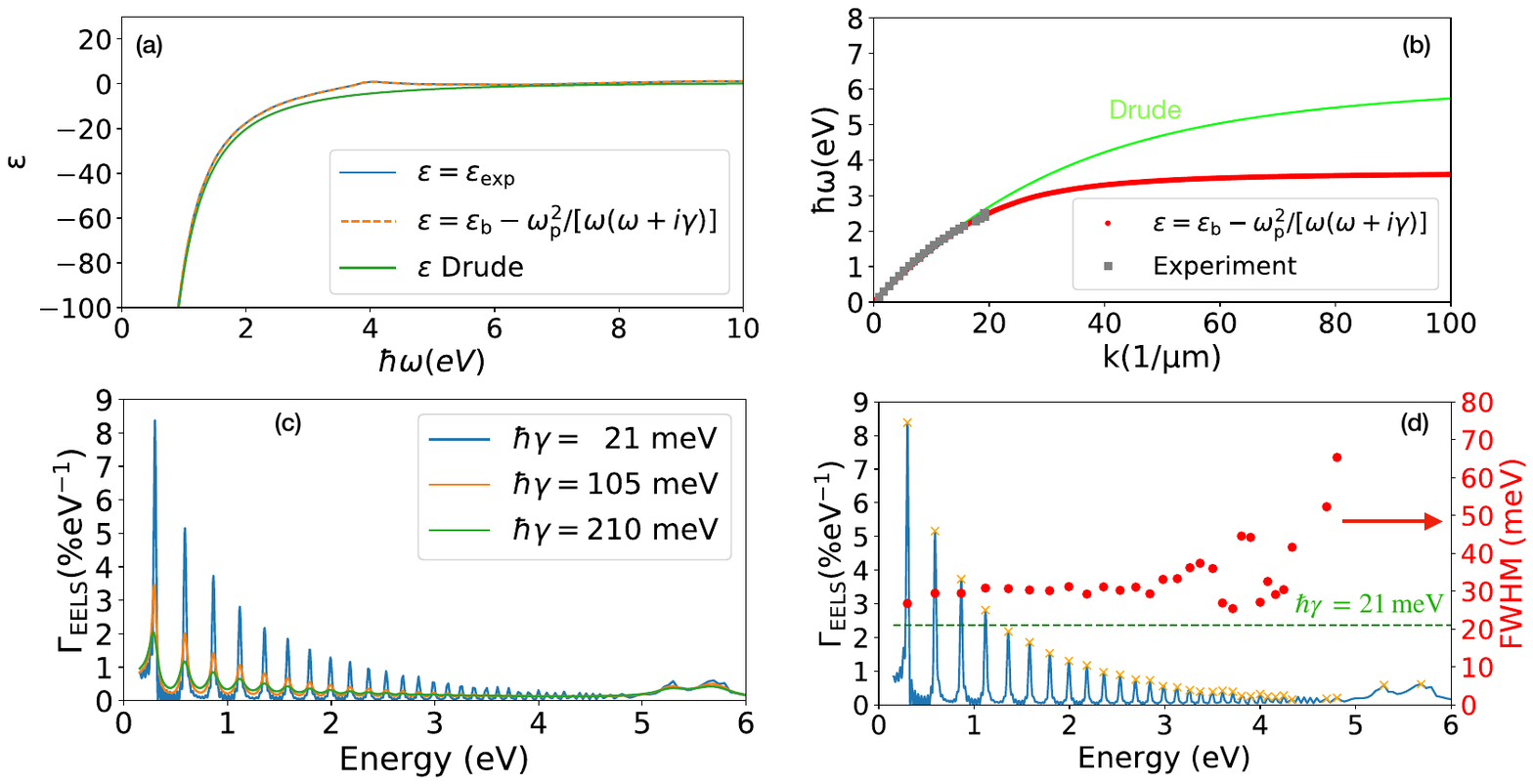}
    \caption{\textbf{Effect of interband transitions in the dispersion relation.} 
    \textbf{a} Real part of the measured dielectric function 
    $\epsilon_{\rm exp}$ of silver \cite{JC1972} and associated 
    Drude models with $\eps_0=1$ and non-vanishing background dielectric
    function (see text for details).
    \textbf{b} Dispersion relations of the fundamental SPP modes 
    using the data/models in (a), and compared to measured values 
    (squares symbols) from experiments, for a nanowire of radius $R=30$~nm.
    \textbf{c} EEL spectra for different damping rates $\gamma$ (see legend).
    \textbf{d} (Simulations) Left axis: EEL spectra for a finite cylinder of length $L=1.2$~$\um$ and radius $R=15$~nm, for an electron trajectory at
    the right edge of the nanowire (i.e., $x_0=L/2$ and $b=5$~nm). Yellow crosses indicate the maximum of the peaks. Right axis: full-width at half maximum (FWHM) of the peaks, where the green-dashed line is fixed at the intrinsic Drude damping rate $\hbar \gamma = 21$~meV obtained by
    fitting to experimental data \cite{JC1972}. 
    }
    \label{fig:S4}
\end{figure}

From measured optical data~\citenum{JC1972}, we obtain the dielectric 
function of silver $\eps_{\rm exp}(\omega)$ which we fit to the Drude model 
\begin{equation}
    \eps(\omega) =\eps_0 - \frac{\wp^2}{\omega(\omega+\ii\gamma)}.
\end{equation}
Here, we use $\hbar\wp=9.17$~eV, $\hbar\gamma=21$~meV, and $\eps_0=1$. 
However, for frequencies $\hbar \omega > 2$~eV, interband transitions 
start to play a role, and therefore, we compute a background permittivity $\eps_{\rm b} = \eps_{\rm exp}+\wp^2 / [\omega(\omega+\ii\gamma)]$ that 
replaces $\eps_0$ and approximately includes the effect of core electrons. 
In Fig.~\ref{fig:S4}a, we display all data sets (i.e., the experimental 
optical data, the Drude with the background permittivity, and the Drude 
model with $\eps_0=1$). 

Interestingly, in Fig.~\ref{fig:S4}b, we show the corresponding dispersion relation of the fundamental SPP modes of an infinite silver nanowire of 
radius $R=30$~nm, and compare it to experimental data.

\newpage
\section{Surface plasmon propagation velocity}
\begin{figure}[ht]
	\centering
	\includegraphics[trim = 0mm 75mm 0mm 35mm, clip, width=1
    \textwidth]{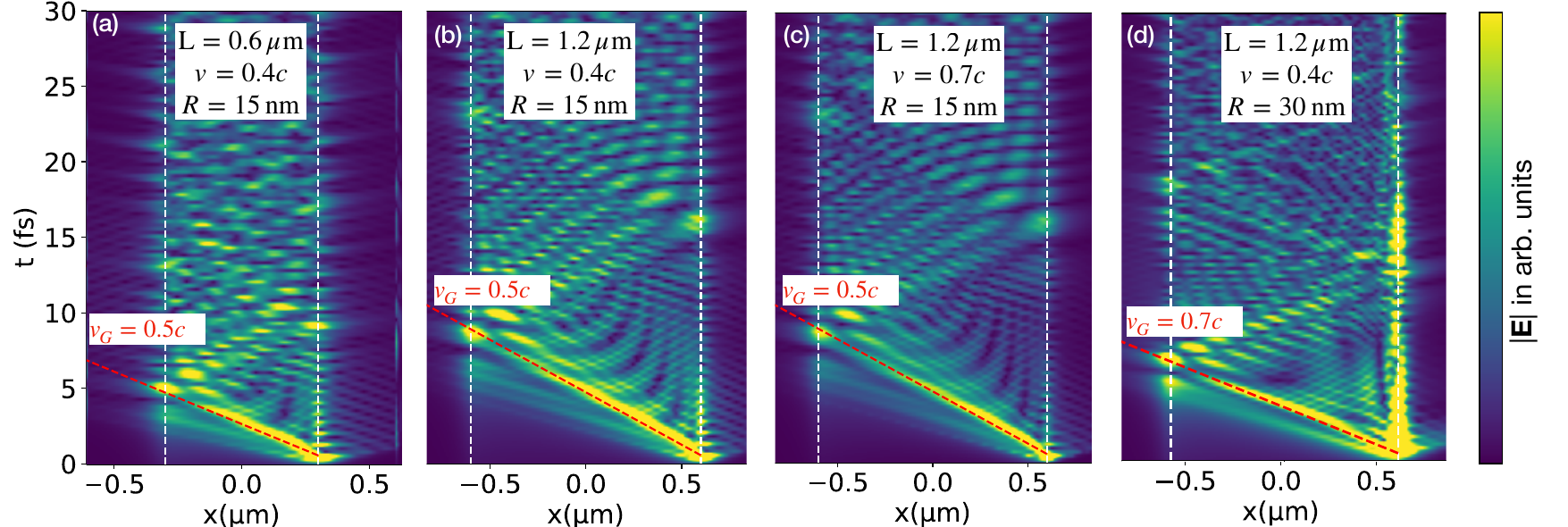}
	\caption{Electric field amplitude along a line along the $x$-axis of  nanowires of \textbf{a} length $\rm 0.6$~$\um$ and textbf{b} length
		$\rm 1.2$~$\um$ for an electron beam of velocity $v=0.4c$ passing 
		aloof at the right edge (i.e., $x_0=L/2$ and $y_0=20$). 
		\textbf{c} same as (b) but with electron velocity $v=0.7c$. 
		\textbf{d} same as (b) but with radius $R=30$~nm. 
		In all panels, the red-dashed line fits the propagation velocity 
		$v_G$ of the fundamental SPP modes that move from right to left.
 }
	\label{fig:S5}
	\end{figure}

In order to determine the group velocity of the fundamental SPP modes, 
we record the electric fields along a line of the $x$-axis near the 
surface of the nanowire. In Fig.~\ref{fig:S5}, we depict the time evolution 
of the field's absolute value $|\vect{E}|$. The group velocities are determined using an excitation via an aloof electron beam near the 
right edge of the nanowire. We observe that at time $t=0$, the SPP modes 
are excited and, after a tiny delay $\sim 1$~fs, propagate towards the 
left. Overall, a bright SPP wave packet that propagates with a well-defined group velocity exists and we obtain the velocity by fitting its slope. 
Behind this primary wave SPP packet feature, there occur several satellites that move to the left at slower velocities (Here, we do not characterize
these satellites quantitatively). For the nanowire with radius $R=15$~nm, 
we obtain for the fundamental SPP wave packet a group velocity $v_G=0.5c$, which is independent of the length of the wire and the electron velocity.
For the nanowire of $R=30$~nm, we obtain a group velocity $v_G=0.7c$.
%

\newpage
\section{DGTD Simulations}
\begin{figure}[b]
\includegraphics[trim = 0mm 125mm 0mm 75mm, clip, width=1.0\textwidth]{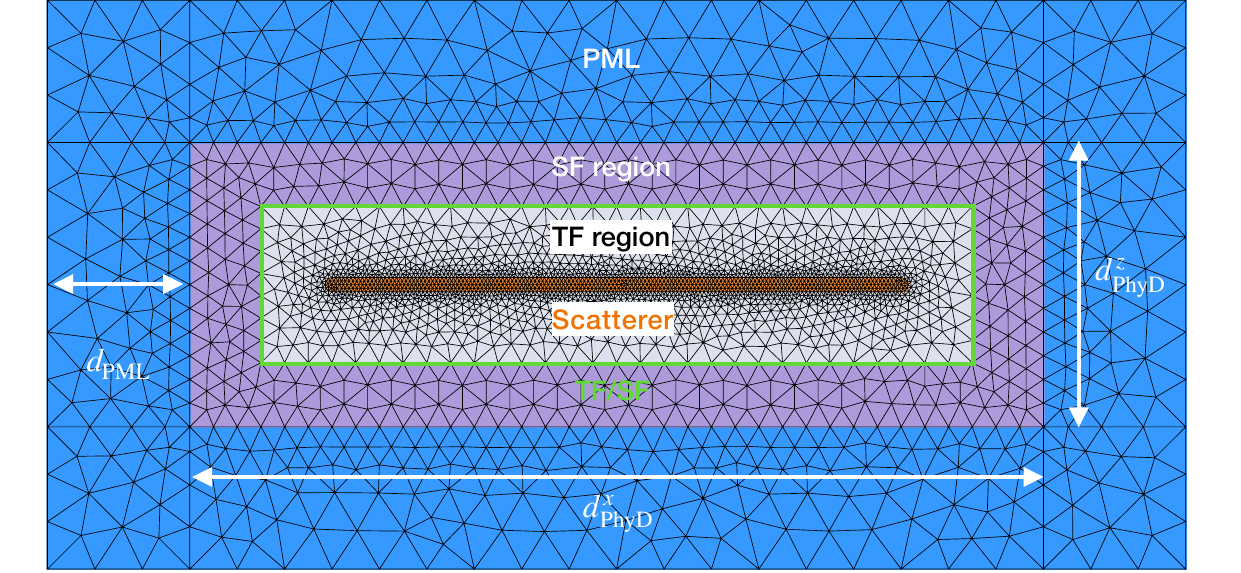}
    \caption{\textbf{Mesh generated by GMSH for DGTD simulations.} Two-dimensional cut along the $xz$-plane through the center of the 
    three-dimenensional mesh for a the simulation of a plasmonic
    nanowire. The mesh has been generated by GMSH and depicts several
    regions used for the numerical computations. From the outermost layer 
    inwards, we identify in sequence: the perfectly matched layer (PML) 
    of thickness $d_\mathrm{PML}$ for realizing absorbing boundary
    conditions, the scattering field 
    (SF) region, the total field (TF) region, and the actual scatterer (i.e., the nanowire). 
    The scatterer, together with the TF and SF region, comprises 
    the physical domain, which is a cuboid of sidelengths 
    $d^x_\mathrm{PhyD}$, $d^y_\mathrm{PhyD}$ and $d^z_\mathrm{PhyD}$. 
    Here, the $y$-direction is normal to the plane of the image.
    }
    \label{fig:S6}
\end{figure}

\subsection{DGTD fundamentals}
The electromagnetic problem in this work consists of calculating the 
induced electric fields that result when a swift electron beam passes
in close proximity (or through) the nanowire. These fields are recorded
and post-processed to yield EEL spectra and entire EELS maps that
characterize the plasmonic excitations. In order to obtain
these induced fields, we solve the full vectorial Maxwell equations
\begin{align}  \label{Eq:Maxwell}
    \partial_t \vect{H} (\vect{r}, t) = &   
    - \mu_0^{-1}\mu^{-1}(\vect{r}) \,  \vecnabla \times \vect{E} (\vect{r}, t) \\
    \partial_t \vect{E} (\vect{r}, t)= &  
    \, \varepsilon_0^{-1}\varepsilon^{-1}(\vect{r}) \, \left[ 
    \vecnabla \times\vect{H} (\vect{r}, t) - \vect{j}(\vect{r}, t)\right],
\end{align}
with the Discontinuous-Galerkin Time-Domain (DGTD) finite-element method \cite{DGTD}, where we assume a linear and isotropic relative permittivity $\varepsilon(\vect{r})$ and a  relative permeability $\mu(\vect{r}) \equiv 1$ at each point in the computational domain. The current density $\vect{j}(\vect{r}, t)$ is used to model dispersive materials via the
technique of auxiliary differential equations\cite{DGTD}.

As a first step, we create an unstructured mesh of the computational domain using tetrahedral elements. This is accomplished with the open-source 
finite-element mesh generator GMSH \cite{geuzaine2009gmsh} (see Fig.~\ref{fig:S6}), where the simulation domain is divided into four 
regions: the scatterer (i.e., the nanowire), the total field (TF) region enclosed by a total field/scattered field contour (TF/SF-contour), scattered field (SF) region and the perfectly matched layer (PML). The latter is
used to realize an reflectionless and absorbing boundary region, thus
emulating an infinite computational domain. The outer boundaries of the simulation domain are terminated by a Silver-Müller outgoing wave
boundary condition that minimizes potential spurious reflections \cite{DGTD}. Within this method, the mesh size is adjusted as required for faithfuly
representing the geometrical details of different regions. As displayed
in Fig.~\ref{fig:S6}, the scatterer region is tessellated very finely 
since the electromagnetic fields in the proximity of the nanowire 
exhibit strong spatial variations, while the domain far from the
scatterer features larger elements in order to reduce memory usage
and to save simulation time.

In the simulation, the electromagnetic fields in each element are
expanded in a polynomial basis using Lagrange polynomials of adjustable
order. Across element boundaries, the appropriate interface conditions
are enforced through a so-called numerical flux, where the distinct field values at either side of a boundary, which result from separate solutions of Maxwell equations, are replaced by a weighted average obtained from the analytical solution of the Riemann problem (upwind flux) \cite{DGTD}. 
For time-stepping, a low-storage Runge-Kutta scheme is employed.

For the implementing of an the electron source,
we utilize the scattered-field source (SF-source) approach, where the polarization current  that is induced in the volume of any non-vacuum 
material  by an electron moving at constant velocity along a 
prescribed trajectory is (analytically) determined and added to the 
current density distribution in Eq.~\eqref{Eq:Maxwell}. To avoid 
numerical artifacts and in order to simulate more realistically the experimental situation that involve electron beams of finite width, 
we model the electron beam with a Gaussian charge distribution of the 
form
\begin{equation}   \label{eq:Gaussiancharge}
    \rho(\vect{r}) = -e \frac{ \mathrm{e}^{-r^2/\sigma_e^2}}{\sigma_e^3 \sqrt{\pi^3} },
\end{equation}
which fulfills the normalization condition $\int d^3\vect{r}\,  \rho(\vect{r}) = -e$. The corresponding Gaussian width is assigned the value 
$\sigma_e = 5$\,nm \cite{Elli_Wenhua2023}, so that it is smaller 
than the size of any other characteristic length in the system.

\subsection{EELS computations}
\begin{figure}[b]
\includegraphics[trim = 0mm 42mm 10mm 10mm, clip, width=1
    \textwidth]{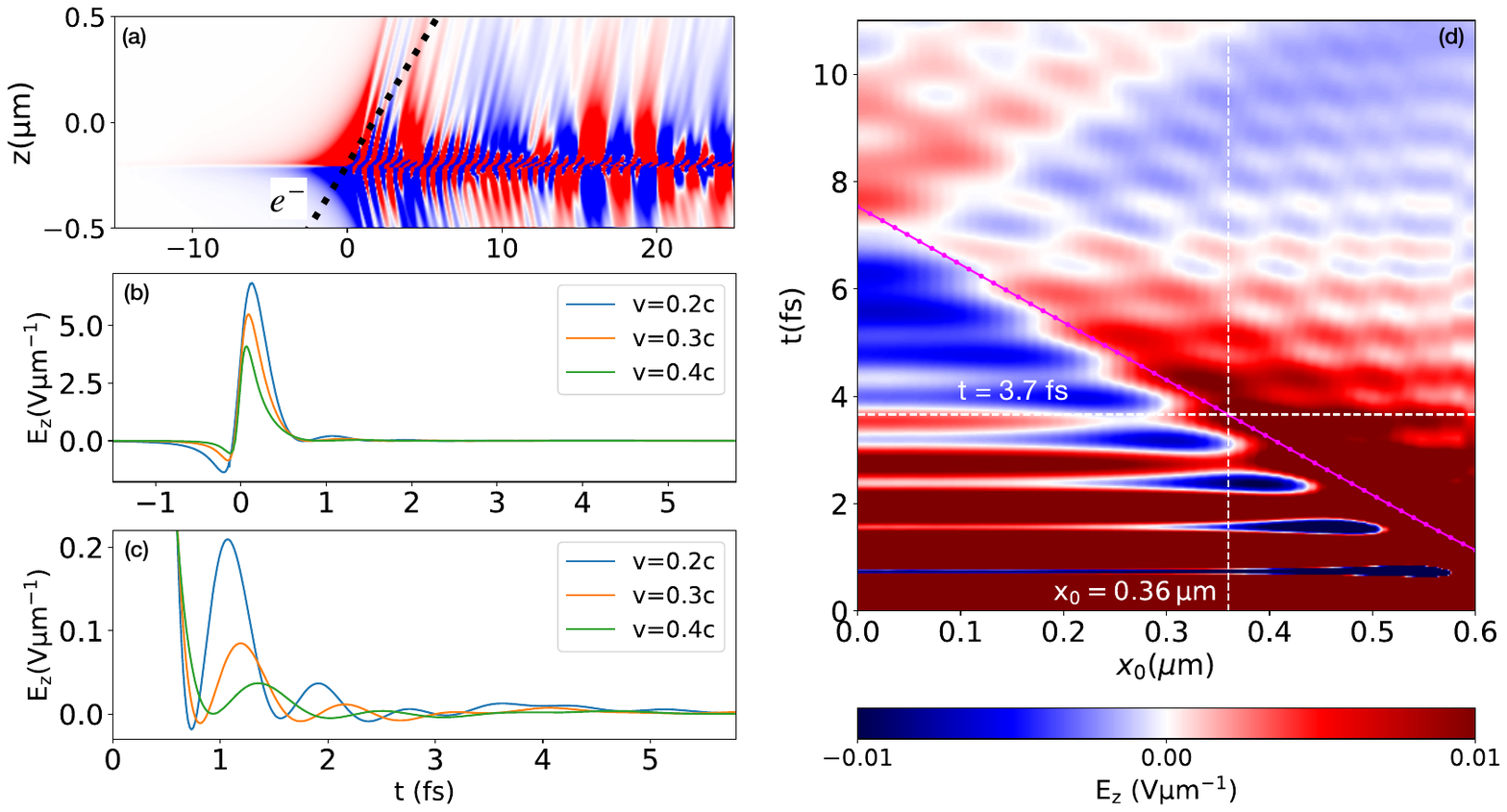}
    \caption{{\bf Electric fields acting on the electron.} \textbf{a} Induced electric field $E_{z}^\ind$ in the interaction of the nanowire with a fast electron beam traveling with velocity $v=0.4 c$ (kinetic energy $\approx 47$~keV) at $x_0=0.36$ ~$\um$, evaluated along the trajectory of the electron on the $z$ axis as a function of time. The black dotted lines indicate the position of the electron in space and time. We only show a time window of 40\,fs, while the total simulation time is $200$\,fs. We note that the electron arrives at the position of closest proximity to the nanowire  around $t\approx0$\,fs. \textbf{b} Electric field at the position of the electron for different electron velocities (see legend). \textbf{c} Zoomed version of (b) between $0$\,fs and $5$\,fs. \textbf{d} Electric field on the electron position for different trajectories along $x$ axis with electron velocity $v=0.2c$, together with return time $t_R=t_1+t_2$ in pink color (see Eq.~(5) in the main text).}
    \label{fig:S7}
\end{figure}

When solving the Maxwell equations~\eqref{Eq:Maxwell} as described in the 
above subsection, we record the induced electric field both in time and space,
either along the trajectory of the electron or on/in any other desired  planes/surfaces/volumes. For a swift electron, we employ the non-recoil approximation, where the electron travels in a straight line parallel to 
the $z$-axis with constant velocity $v$. Therefore, its position on the 
$z$-axis follows the trajectory $z(t) = vt+z_0$, with $z_0$ being the $z$ component of its initial position. The electrons energy loss $\Delta E$ 
when passing by the nanowire is then expressed as 
$\Delta E = \int_0^\infty \hbar\omega \, d \omega \, \Gamma_\EELS(\omega)$, 
where
\begin{equation}
    \Gamma_\mathrm{\EELS} (\omega) = \frac{e}{\pi\hbar\omega} \int dz\, \real{ \ee^{-\ii\omega \frac{z-z_0}{v}} E_{z}^\ind (z, \omega) }
\end{equation}
denotes the probability that the electron loses an energy of the amount
$\hbar \omega$.
Further, $E_{z}^\ind (z, \omega) $ is the Fourier transform of the 
$z$-component of the induced electric fields $E_{z}^\ind (z,t) = \uvec{z} \cdot \Eb^\ind$, recorded directly by the DGTD computations along the electron
trajectory, as shown in Fig.~\ref{fig:S7}a (see dotted line) where we 
depict the $z$-component of the induced electric field $E_{z}^\ind (z,t)$ 
by an electron moving at velocity $\vb = 0.4 c \,\uvec{z}$ for an
electron trajectory at $x_0=0.36\, \rm \mu m$. The black-dotted line 
indicates the position of the electron (in space and time) when it
traverses the computational domain. When moving along this trajectory,
the electron experiences an accelerating/decelerating force. In other
words, the induced field does work against the electron. However, we
would like to highlight the change of sign in blue and red of the 
electric field at the position of the electron along the trajectory.
This indicates that while the overall effect of the induced electric
field acting on the electron leads to a net energy loss, there are
positions/times on the trajectory where the electron experiences
differential gain. The electric field on the position of the electron for different velocities is displayed in Fig.~\ref{fig:S7}b, with a zoomed-in version in Fig.~\ref{fig:S7}c. In Fig.~\ref{fig:S7}d, we plot the electric 
field on the position of the electron for different trajectories 
along the wire for an electron beam of velocity $0.2c$. This should
be compared with the return time $t_R$ in pink color for the plasmon modes to return to the electron, see Eq.~(5) in the main text.

\subsection{Parameters used in the DGTD simulation}

To ensure convergence of the computed EEL spectra, it is necessary to 
optimize a number of parameters in the DGTD simulations. In particular, 
the size of the physical domain is set to be a cuboid of side 
$d^x_{\rm PhyD} = 1.8\, \um$, $d^y_{\rm PhyD} = 0.6\, \um$ and 
$d^z_{\rm PhyD} = 0.6\, \um$. 
We ensure that the field strength in 
the middle of computation domain (i.e., the position of the nanowire) is at least three orders of magnitude larger than that at the boundary of the physical domain, to make sure that we do not disregard important contributions in the EELS. 
The PML thickness is set to $d_{\rm PML} = 0.3\, \um$.
%
%
The characteristic side length of the tetrahedral meshing elements is 
chosen at $h_{\rm max} = 5$\,nm in the scatterer region to ensure that,
when using third-order basis functions on each element, we have the 
required accuracy to resolve the Gaussian charge distribution 
($\sigma_e=5$\,nm) that represents the electron-beam source.
In the far-field region, the element size is relaxed to $h_{\rm max} = 50-100$\,nm. The time-stepping ($\approx 0.003$~fs) is done by a 
fourth-order low-storage Runge-Kutta integrator~\cite{DGTD}. 
The total physical simulation time is set to $200\,$fs.

\newpage


\section{Guided modes in the substrate}

\begin{figure*}[b]
    \centering
    \includegraphics[trim = 0mm 58mm 0mm 85mm, clip, width=1\textwidth]{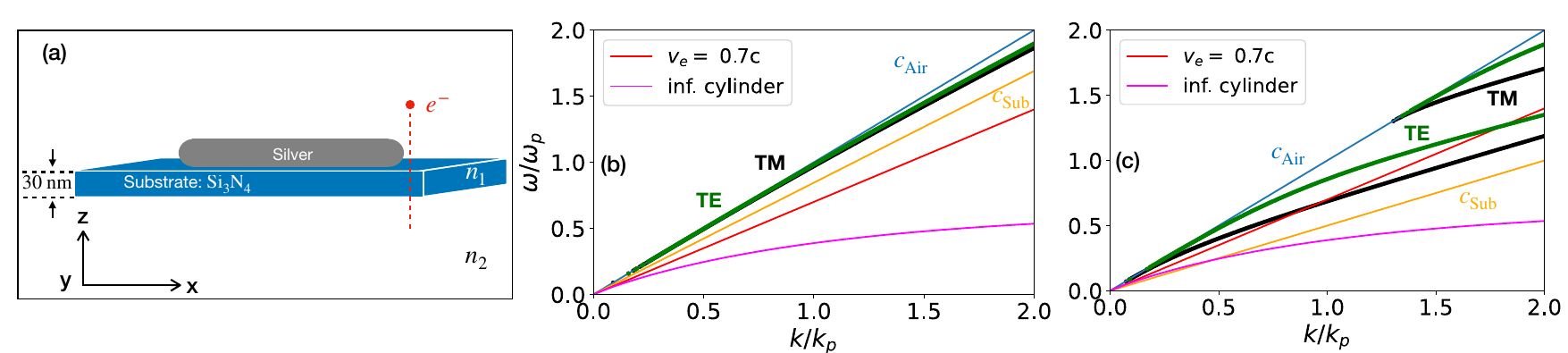}
    \caption{\textbf{Effect of the substrate.} 
    \textbf{a} Schematic for a nanowire on an amorphous silicon nitride substrate 
    with an electron beam passing in close proximity to the nanowire 
    in aloof configuration. 
    \textbf{b} Dispersion relations of substrate-guided modes, light
    cones in vacuum and material, and electron lines for a low-permittivity
    substrate with $\eps_{\rm sub}=1.4$.
    \textbf{c} same as b, for a high-permittivity substrate with 
    $\eps_{\rm sub}=4$. See text for further details.
} 
\label{fig:S8}
\end{figure*}


In the experiments, the silver nanowires are deposited on a 30~nm-thick silicon nitride substrate with permittivity $\eps_{\rm sub} \approx 4$ in the visible frequency range \cite{plasmon_waveguiding, Luke2015_optical_constants}
(see  Fig.~\ref{fig:S8}a). 
Such thin films could sustain guided modes that can be excited with the electron beam. Therefore, to understand in more detail the influence of the substrate, we plot in Figs.~\ref{fig:S8}b and ~\ref{fig:S8}c, respectively, 
the analytical TE- (green curves) and TM- (black curves) guided modes of an planar slab of thickness $d_{\rm sub} = 30$~nm with $\eps_{\rm sub}=1.4$ 
and $\eps_{\rm sub}=4$ (see details in the section below). In these 
dispersion diagrams, we also incorporate the electron line (moving at $v=0.7c$), and the light lines of vacuum and of the corresponding slab material. Incidentally, the only qualitative difference between Figs.~\ref{fig:S8}b and ~\ref{fig:S8}c is that the value of the dielectric permittivity of the
substrate leads to the electron line being inside the light cone of 
the material or not. Consequently, in the high-permittivity case, 
the electron line intersects the substrate-guided TE and TM mode 
dispersion curves while there is no intersection for the low-permittivity 
case. In other words, in the high-permittivity case, we expect guided 
modes in the substrate to be excited by the electron beam, while in 
low-permittivity case, we do not expect that guided modes in the 
substrate will be excited.

\begin{figure} 
	\centering
	\includegraphics[trim = 0mm 95mm 0mm 30mm, clip, width=\textwidth]{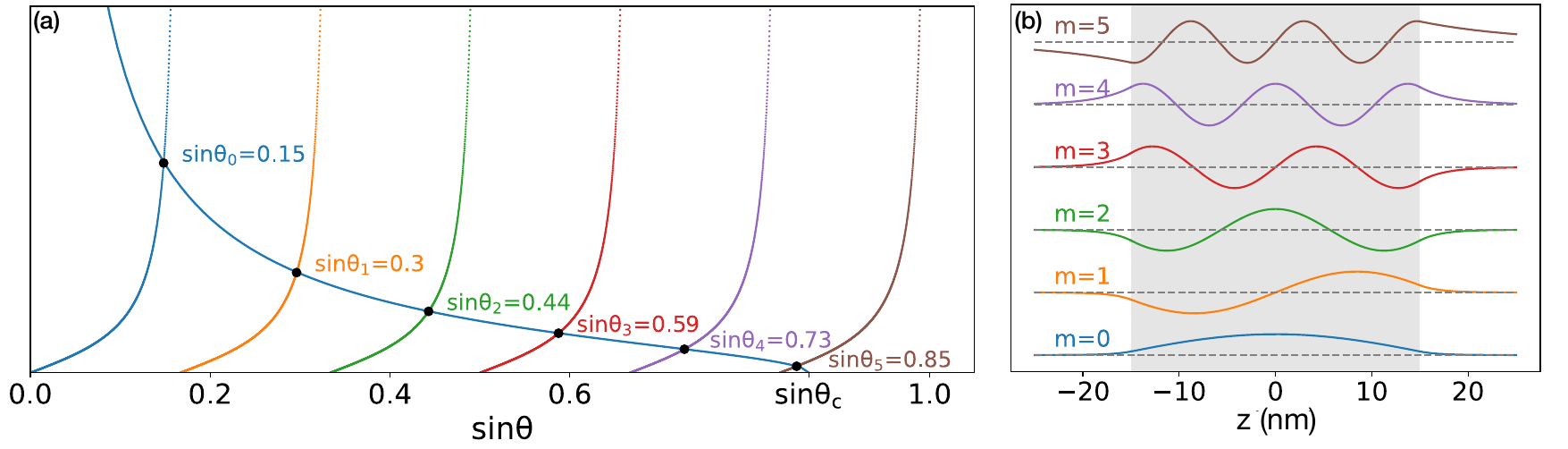}
	\caption{\textbf{a} Graphical solution of Eq.~\eqref{eq:modes_angle_transcendental_TE} for  $\lambda_0 = 20$~nm and $d_{\rm sub} = 30$~nm. \textbf{b} The first five TE modes, for which we show the amplitude of $u_m(z)$ (arb. units).
 } 
	\label{fig:S9}
	\end{figure}

Specifically, we consider a planar dielectric waveguide of thickness 
$d_{\rm sub} = 30$~nm, with a refraction index $n_1=2$ surrounded by 
air with $n_2=1$ (i.e., corresponding to the substrate slab made of 
silicon nitride). Following the approach in Ref. \cite{SalehTeich_Fundamentals}, the critical angle $\overline{\theta}_c$ for total internal reflection is determined by $\rm cos(\overline{\theta}_c)=n_2/n_1=0.5$. 
To obtain the angles $\rm \theta$ of the guided TE modes, we solve the transcendental equation \cite{SalehTeich_Fundamentals}
\begin{equation} \label{eq:modes_angle_transcendental_TE}
    \tan \biggl( \pi \frac{d_{\rm sub}}{\lambda} \sin\theta -m \frac{\pi}{2} \biggr)= \sqrt{\frac{\sin^2\overline{\theta}_c}{\sin^2\theta}-1},
\end{equation}
where $\lambda = \lambda_0/n_1$ denotes the wavelength in the substrate of refective index $n_1 = \sqrt{\epsilon_{\rm sub}}$ and $m$ is the order of the TE-mode. In Fig.~\ref{fig:S9}a we depict the graphical solution of q.~\eqref{eq:modes_angle_transcendental_TE} for $\lambda_0 = 20$~nm and $d_{\rm sub} = 30$~nm, where the intersection points give the desired angles $\theta_m$ of the guided modes.  Incidentally, for TE modes, the electric field is orthogonal to the plane of incidence (i.e., $z$-polarized), and is given by $E_y(x,z) = a_m u_m(z) {\rm exp}(- \ii \beta_m x)$, where the $x$-axis is the propagation direction of the modes. By considering the boundary conditions \cite{SalehTeich_Fundamentals} between substrate and air, we plot $u_m(z)$ in Fig.~\ref{fig:S9}b.

Similarly, for the TM modes but we now solve \cite{SalehTeich_Fundamentals}
\begin{equation} \label{eq:modes_angle_transcendental_TM}
    \tan \biggl( \pi \frac{d_{\rm sub}}{\lambda} \sin\theta -m \frac{\pi}{2} \biggr)= \frac{\sqrt{1-\sin^2\theta-cos^2\overline{\theta}_c}}{\sin\theta \cos^2\overline{\theta}_c}.
\end{equation}
Once we obtain $\theta_m$, we calculate the $x$-component of the guided wavevectors 
in the substrate as
\begin{equation} \label{eq:modes_betas}
    \beta_m = n_1 k_0 \cos \theta_m,
\end{equation}
where $k_0 = 2 \pi/\lambda_0$.
The above-determined guided modes of the substrate can be excited by the swift electrons used to launch SPPs in the nanowire, with the electrons also passing through the substrate. The analysis of the guided modes excited in the substrate proceeds along similar lines as the study of the excitation of the SPPs. However, the efficiency of this process is considerably weaker than the plasmonic processes.

\section{Videos}

The file "Movie$\_$EXP.mp4" shows the temporal evolution of the experimental excitation probability in the $xy$ plane, via a temporal Fourier transform of the experimental EEL spectra. The experimental setup is the same as in Fig.~2 of the main text, with the difference that here we scan the entire $xy$ plane with electron beams.

The file "Movie$\_$DGTD.mp4" shows the temporal evolution of the amplitude of the electric field $|\Eb(\rb,t)|=\sqrt{E_x^2+E_y^2+E_z^2}$ in the $xz$ plane. The position of the electron is indicated as a red point following the $z$ axis for a specific electron trajectory with $x_0=0.36$~$\um$ and impact parameter $b=5$~nm. The electron passes by the center of the freestanding nanowire around time $t\approx0$~fs. The nanowire is $1.2$~$\um$ long and has a diameter of 30~nm that is modeled with a Drude permittivity: plasma frequency $\hbar \wp = 9.17$~eV, damping $\hbar \gamma = 21$~meV, and $\eps_0=1$.

%